# Investigating the Correlation between Force Output, Strains, and Pressure for Active Skeletal Muscle Contractions

Karan Taneja[a], Xiaolong He[b], Chung-Hao Lee[c], John Hodgson[d], Usha Sinha[e], Shantanu Sinha[f], J. S. Chen[a],*

a Department of Structural Engineering, University of California San Diego, La Jolla, CA, USA

b ANSYS Inc, Livermore, CA, USA

c Department of Bioengineering, University of California Riverside, Riverside, CA, USA

d Department of Integrative Biology and Physiology, University of California Los Angeles, Los Angeles, CA, USA

e Department of Physics, San Diego State University, CA, USA

f Department of Radiology, University of California San Diego, La Jolla, CA, USA

**Abstract**

Measuring the forces of individual muscles in a muscle group around a joint is non-trivial, and researchers have suggested using surrogates for individual muscle forces instead. Traditionally, experimentalists have shown that the force output of the skeletal muscle tissue can be correlated to the intra-muscular pressure (IMP) generated by the muscle belly. However, IMP proves difficult to measure *in vivo*, due to variations from sensor placement and invasiveness of the procedure. Numerical biomechanical simulations offer a tool to analyze muscle contractions, enabling new insights into the correlations among non-invasive experimentally measurable quantities such as strains, and the force output. In this work, we investigate the correlations between the muscle force output, the principal, shear and volumetric strains experienced by the muscle, as well as the pressure developed within the muscle belly as the tissue undergoes isometric contractions with varying activation profiles and magnitudes. It is observed that pressure does not correlate well with force output under higher sub-maximal and maximal activation levels, especially at locations away from the center of the muscle belly due to pressure relaxation effects. This study reveals strong correlations between force output and the strains at all locations of the belly, irrespective of the type of activation considered. This observation offers evidence for further *in vivo* studies using experimentally measurable principal and volumetric strains in the muscle belly as proxies for the force generation by the individual muscle and consequently enables the estimation on the contribution of various muscle groups to the total force.

## 1 Introduction

Skeletal muscle fibers generate active and passive forces through the contraction of their constituent sarcomeres, where cross-bridges form between the actin and myosin filaments modulated via adenosine triphosphate (ATP) hydrolysis and calcium ions [1]. A bundle of these fibers surrounded by the passive endomysium constitute a muscle group [2,3], and multiple muscle groups working together produce the total force that acts on a joint, resulting in its movement. One of the major challenges in muscle physiology is to understand the contribution of individual muscle groups to the total force generated and thereby, the overall joint torque – a metric related to risk of injuries. Although biomechanical modeling has provided a means to analyze muscle groups under external loading [1,4], it currently lacks the capability to directly relate the stress produced by a single contracting muscle to the overall mechanical performance of the muscle groups.

One way to overcome this challenge is to use surrogate measures of muscle stress relative to the maximum voluntary contraction (MVC) of a muscle [5], where MVC is the maximum force a muscle group can produce under a controlled condition (e.g., an isometric contraction), providing an estimate of the muscle's capacity. However, since measuring the internal force of a muscle during MVC is non-trivial, physiologists resort to other indicators of muscle activity that act as surrogate measures, including muscle electrical activity through surface electromyography (sEMG) and the intra-muscular pressure (IMP) from a muscle contraction [6,7]. Here, sEMG is a measure of the electrical activity of the muscle as it contracts, which can be measured by placing non-invasive electrodes directly on the skin. While it is not a direct measure of force, higher sEMG signals correspond to higher muscle activity [8]. One disadvantage of this modality is that the amplitude of sEMG does not change proportionately to the modification in a muscle's force [9], but is affected by logistical constraints such as the accuracy of electrode placement and influence of subcutaneous tissue [10]. Moreover, the relationship between muscle activity and force is not linear, with force production also depending on the relationships between muscle tension, muscle length, and contraction velocity [1]. Thus, it may not be feasible to directly use sEMG as a surrogate measure of a single muscle's stress.

On the other hand, IMP is the hydrostatic pressure of the muscle's interstitial fluid [6,11] generated from a contracting muscle group that can serve as a direct measure of the mechanical state of a muscle. *Ex vivo* studies of isolated anterior tibialis muscle across different species have shown that IMP is fairly well correlated to muscle stress under static conditions [12–14]. More recently, using a minimally invasive approach based on fiber optic sensors, Ateş and coworkers [6,11] have shown *in vivo* that IMP correlates with the mechanical performance of individual muscles for young and old adults. However, it should be noted that IMP is a measure of pressure in a *localized* region and varies with the depth and geometry of a muscle. Moreover, it does still require an invasive procedure, albeit minimally, and a completely non-invasive method would be preferable for studying normal and diseased muscle functions.

The present work is motivated to explore the correlation of other surrogates, which are measurable in a clinical imaging environment and can be conveniently applied *in vivo*, to estimate muscle force generation non-invasively. A useful recent addition to the experimental data on muscle physiology has been the use of magnetic resonance imaging (MRI) in human subjects to provide higher-resolution data on intramuscular deformation during muscle activity [15]. For example, MRI techniques were used by Jensen *et al.* [16,17] to investigate the regional variation in volumetric strain. They proposed a relationship between IMP and volumetric strain and suggested that IMP sensors should be placed in the relatively homogeneous areas of volumetric strain to produce reliable force measurement surrogates. Other MR-based studies have also examined the correlations between strains and force output of the muscle. For example, Malis *et al.* [18,19] indicated a significant role of principal and shear strains during skeletal muscle contraction for force production, which were strong indicators of force variations amongst young and old subjects [20,21].

While these studies independently establish either direct or indirect relationships between different surrogate measures and force, a direct correlation between them and individual muscle force has not been established yet. A computational model can examine these relationships between IMP, strains, and individual muscle force output. Establishing these correlations will enable *experimentally observable* deformation indices (e.g., shear/volumetric strain) via non-invasive MRI to be used as *in vivo* surrogate markers of individual muscle force. Further, such a correlation study among these quantities at different spatial locations of the muscle will provide physiologists with useful guidance on the regions of interest to perform imaging as well as on the accurate placement of sensors to estimate IMP.

Numerical simulations have been employed to investigate the behaviors and mechanisms of skeletal muscle tissues. Starting from initial simplistic one-dimensional lumped parameter models [22], multi-dimensional continuum models have been favored by researchers in recent times [23–26]. The anisotropic properties of skeletal muscle have been simulated using transversely isotropic hyperelastic material models [23–25] with anisotropy introduced, e.g., through modeling the effect of collagen fibers wrapped around the muscle fiber [23]. However, these models do not account for the viscous effects observed for skeletal muscle tissues in mechanical characterization experiments [27]. It is to be noted that viscoelasticity plays an integral role in the development of IMP during muscle contraction over time, resulting from the active and passive responses of a 3-D composite of muscle fibers and the extracellular matrix (MT) [27,28], inducing damping in the muscle contractions. To generate physiologically relevant computational models, researchers have included visco-hyperelasticity to model skeletal muscle behavior [28–30]. To that end, several studies have also incorporated viscous effects to explore the IMP-force relationship. For example, Bojairami and Driscoll [31] correlated IMP and force via a shell finite element model by considering a neo-Hookean strain energy function for the skeletal muscle. Wheatley *et al.* [32] calibrated a model to explore fluid content within the muscle tissue to predict force and IMP, further elucidating the mechanisms of IMP spatial variability in the muscle [33]. Despite these model developments, a comprehensive analysis of the relationships between the force, IMP, and deformation metrics that account for sophisticated hyperelastic and viscous muscle behaviors has not yet been conducted.

The present study describes our initial steps to develop a framework to explore the relationship between IMP, mechanical strains (principal, shear, volumetric), and force output. We use a visco-hyperelastic material model to simulate the isometric contractions for sub-maximal and maximal linear and non-linear muscle activation profiles. Through a systematic parametric study, we identify the correlations between the force output of the muscle model, the pressure as well as various strain measures calculated at different locations in the muscle belly. This allows one, *for the first time*, to derive meaningful conclusions on the relationship of observable deformation outputs, specifically the strains, and its influence on muscle force generation during contractions. This observation can enable an estimation of the contribution of various muscle groups to the total force, by experimentally measurable principal and volumetric strains in the muscle belly. The remainder of this manuscript is organized as follows. Section 2 provides an overview of the material models of various components in the continuum-scale muscle model. In Section 3, the continuum-scale skeletal muscle simulations for isometric contractions are performed to systematically examine the influences of activations on muscle force output, pressure, and mechanical strains. Finally, concluding remarks are made in Section 4.

## 2 Methods

### 2.1 *Material models for the continuum muscle-tendon complex*

The continuum muscle-tendon complex (Figure 1(a)) contains muscle fibers, anisotropic matrix, aponeurosis, and tendon. The following subsections describe the employed hyperelastic and visco-hyperelastic material models.

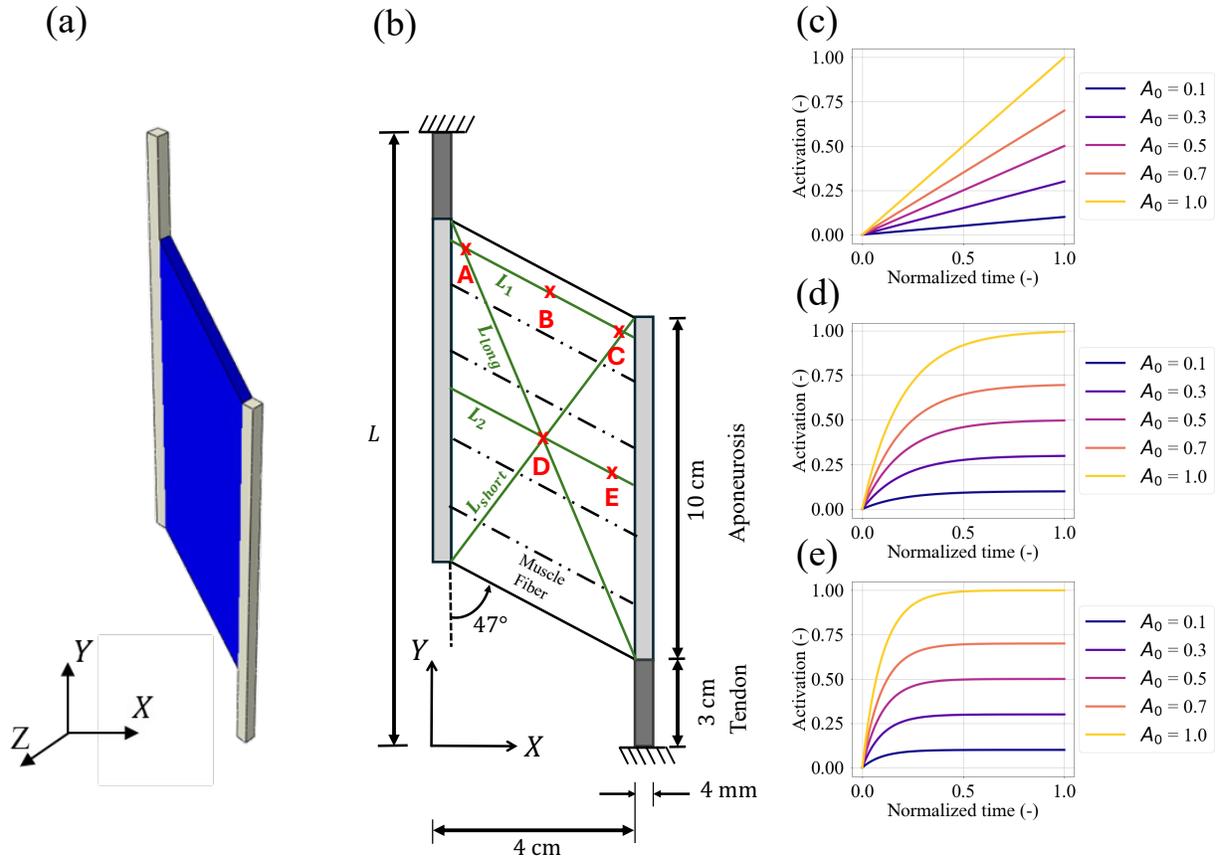

Figure 1: An overview of the proposed modeling of the skeletal muscle. The model in (a) is a continuum-scale plane strain model with an initial pennation angle of 47° and a thickness of 0.4 cm in the $Z$-direction, where the dash-dot lines represent the orientation of muscle fibers in (b). The lines $L_1, L_2$ along the direction of the muscle fibers, and $L_{long}, L_{short}$ along the diagonals across the belly are the directions along which distributions of pressure and strains are examined. The five positions $A \rightarrow E$ in the belly indicate the locations where the pressure and strains are extracted for the statistical analysis. The plots in (c-e) show the fifteen different activation profiles described in Table 4.

*2.1.1 Hyperelastic response of tendon and aponeurosis*

In the continuum muscle-tendon complex, the tendon and the aponeurosis are modeled by an isotropic third order generalized Rivlin model to represent the softer responses in the low-strain region and the stiffer responses in the high-strain region, given as

$$W_{\text{tendon}}(\bar{I}_1, J) = \sum_{i=1}^{3} C_{i0}(\bar{I}_1 - 3)^i + \frac{K_{\text{tendon}}}{2}(J-1)^2. \tag{1}$$

Herein, $\bar{I}_1 = J^{-2/3} I_1$, are the reduced invariants of the right Cauchy-Green tensor $\mathbf{C} = \mathbf{F}^T \mathbf{F}$ with $I_1 = tr(\mathbf{C})$, $I_2 = \frac{1}{2}[I_1^2 - tr(\mathbf{C}^2)]$, $J = \det(\mathbf{F})$, $\mathbf{F} = \frac{\partial \mathbf{x}}{\partial \mathbf{X}}$ is the deformation gradient, where $\mathbf{x}$ and $\mathbf{X}$ are the deformed and undeformed material coordinates, respectively. The material constants given in Table 1 are considered from previous studies [23,34,35].

Table 1: Material parameters of $W_{\text{tendon}}$ in Eq. (1) for tendon and aponeurosis (unit: N/cm²).

| $C_{10}$ | $C_{20}$ | $C_{30}$ | $K_{\text{tendon}}$ |
|---|---|---|---|
| 30 | 80 | 800 | $10^5$ |

*2.1.2 Visco-hyperelastic response of the skeletal muscle*

As the continuum skeletal muscle contains both fiber and matrix components, the hyperelastic strain energy density function of the continuum skeletal muscle is decomposed into three parts, (i) a passive deviatoric matrix (MT), (ii) a passive volumetric MT, and (iii) an anisotropic (contractile) portion of muscle fibers (FB):

$$W_{\text{muscle}} = W_{\text{MT}}^{\text{dev}}(\bar{I}_1, \bar{I}_2, \bar{I}_4) + W_{\text{MT}}^{\text{vol}}(J) + W_{\text{FB}}^{\text{ani}}(\lambda, \dot{\lambda}), \tag{2}$$

where the deviatoric part $W_{MT}^{dev}(\bar{I}_1, \bar{I}_2)$ is defined as a generalized Rivlin model to describe the transversely isotropic passive material properties:

$$W_{MT}^{dev}(\bar{I}_1, \bar{I}_2, \bar{I}_4) = \sum_{i+j=1}^{3} C_{ij}(\bar{I}_1 - 3)^i(\bar{I}_2 - 3)^j + k_0\{\exp[k_1(\bar{I}_4 - 1)^2] - 1\}, \quad (3)$$

with $\bar{I}_2 = J^{-4/3} I_2$ and an exponential term of the stretch ratio associated with muscle fibers and collagen fibers along the fiber direction, $\bar{I}_4 = J^{-2/3} \mathbf{N} \cdot \mathbf{C} \cdot \mathbf{N}$, where $\mathbf{N}$ is the unit vector in the fiber direction. The material constants $C_{ij}$ shown in Table 2 were obtained from our previous study [23], where the constants were calibrated to the stress-strain synthetic data from a homogenization protocols on a microstructural representative volume element, that was made to undergo extensive virtual uniaxial, biaxial and shear protocols. The other constant $k_1$ is 0.69 [23]. The volumetric part $W_{MT}^{vol}(J)$ is used to represent nearly incompressible materials,

$$W_{MT}^{vol}(J) = \frac{K_{MT}}{2} \ln(J)^2. \quad (4)$$

with the same bulk modulus $K_{tendon} = K_{MT} = 10^5$ N/cm² used for both $W_{MT}^{vol}$ and $W_{tendon}$.

Table 2: Material parameters of $W_{MT}^{dev}$ in Eq. (3) (unit: N/cm²).

| $C_{10}$ | $C_{01}$ | $C_{20}$ | $C_{11}$ | $C_{02}$ | $C_{30}$ | $C_{21}$ | $C_{12}$ | $C_{03}$ | $k_0$ |
|---|---|---|---|---|---|---|---|---|---|
| 2.23 | -1.03 | -10.89 | 24.13 | -4.89 | 10.65 | -16.36 | 8.21 | -1.30 | 0.58 |

The anisotropic part $W_{FB}^{ani}$ is used to describe the contractile Cauchy stress in the fiber $\sigma_{FB}$, including the active-length dependent and velocity dependent effects as considered by [28,36,37],

$$\sigma_{FB} = \lambda \frac{\partial W_{FB}^{ani}(\lambda, \dot{\lambda})}{\partial \lambda} = \sigma_{max} \frac{\lambda}{\lambda_0} \left(a(\bar{t}) f_{active,L} f_{active,V} + f_{passive}\right), \quad (5)$$

where $\sigma_{\max}$ is the fiber's maximum isometric stress, $\bar{t}$ is the normalized time for activation, $\lambda$ is the fiber stretch, $\lambda_0 = 1.4$ is selected as the optimal along-fiber stretch ratio at which the muscle fiber generates maximum force. $f_{\text{active,L}}$ and $f_{\text{passive}}$ are the normalized active-length dependent and passive parts of the muscle fiber force [23], respectively, expressed as,

$$f_{\text{active,L}} = \begin{cases} 9(\lambda^* - 0.4)^2, & \lambda^* \leq 0.6 \\ 1 - 4(1 - \lambda^*)^2, & 0.6 < \lambda^* \leq 1.4 \\ 9(\lambda^* - 1.6)^2, & \lambda^* > 1.4 \end{cases} \qquad (6)$$

$$f_{\text{passive}} = \begin{cases} 0, & \lambda^* \leq 1 \\ \gamma_1[(e^{\gamma_2(\lambda^* - 1)} - 1)], & 1 < \lambda^* \leq 1.4 \\ (\gamma_1\gamma_2 e^{0.4\gamma_2})\lambda^* + \gamma_1[(1 - 1.4\gamma_2)e^{0.4\gamma_2} - 1], & \lambda^* > 1.4 \end{cases} \qquad (7)$$

where $\lambda^*$ is the normalized stretch as $\lambda^* = \lambda/\lambda_0$. The velocity-dependent fiber force, $f_{\text{active,V}}$, is described as, [36],

$$f_{\text{active,V}} = \begin{cases} \dfrac{1 - \left(\dfrac{\dot{\lambda}}{\dot{\lambda}^{\min}}\right)}{1 + k_c\left(\dfrac{\dot{\lambda}}{\dot{\lambda}^{\min}}\right)}, & \dot{\lambda} \leq 0 \\ d - (d-1)\dfrac{1 + \left(\dfrac{\dot{\lambda}}{\dot{\lambda}^{\min}}\right)}{1 - k_c k_e\left(\dfrac{\dot{\lambda}}{\dot{\lambda}^{\min}}\right)}, & \dot{\lambda} > 0 \end{cases} \qquad (8)$$

where $\dot{\lambda}_{\min}$ is the minimum stretch rate of the fiber. The first equation in Eq. (8) describes the concentric phase, where $k_c$ is a dimensionless constant controlling the curvature of the force vs fiber contraction velocity plot. The second equation describes the eccentric phase, where the muscle develops tension as it lengthens, and another dimensionless constant $k_e$ describes the curvature in this phase. The dimensionless constant $d$ is the offset of the eccentric function. Figure 2 shows both length- and velocity-dependent force production from the muscle fiber.

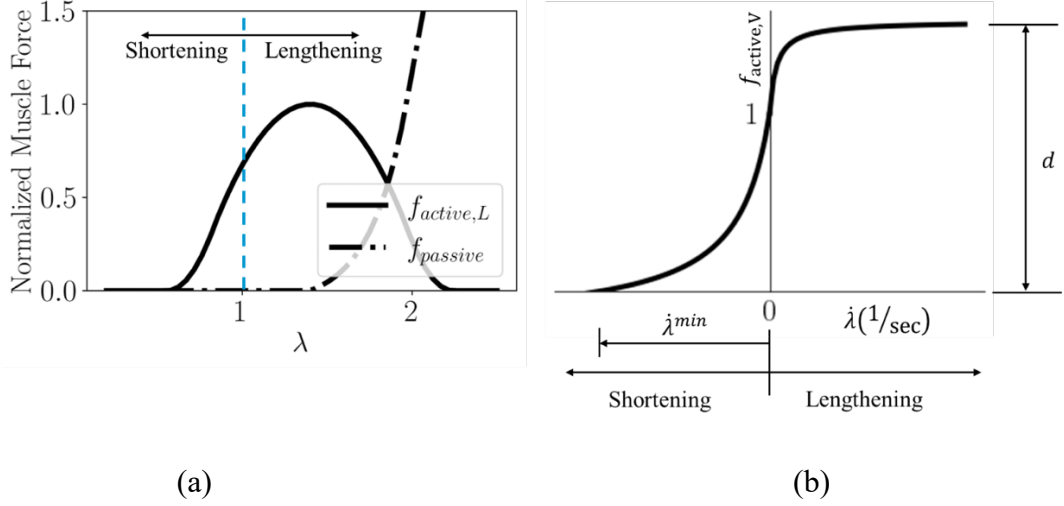

(a)                          (b)

Figure 2: Normalized (a) length-dependent and (b) velocity-dependent mechanical responses of the muscle fiber model with $a(\bar{t}) = 1$. The stretch and stretch rate at which the shortening and lengthening phases of the fiber exist are also indicated.

The viscous effects are only included for the muscle belly. We begin with the hyperelastic formulation of the 2$^{nd}$ Piola-Kirchhoff (PK) stress which can be decomposed into deviatoric, volumetric and muscle fiber stresses,

$$\tilde{S}_{ij} = \tilde{S}_{ij}^{\text{dev}} + \tilde{S}_{ij}^{\text{vol}} + \tilde{S}_{ij}^{\text{FB}}. \tag{9}$$

The deviatoric and volumetric 2$^{nd}$ PK stresses are written as

$$\tilde{S}_{ij}^{\text{dev}} = \frac{\partial W_{\text{MT}}^{\text{dev}}(\bar{I}_1, \bar{I}_2, \bar{I}_4)}{\partial E_{ij}}, \quad \tilde{S}_{ij}^{\text{vol}} = \frac{\partial W_{\text{MT}}^{\text{vol}}(J)}{\partial E_{ij}} \tag{10}$$

where the superscript $'\sim'$ denotes the hyperelastic stresses, and $E_{ij} = \frac{1}{2}(C_{ij} - \delta_{ij}) = \frac{1}{2}(F_{ki}F_{kj} - \delta_{ij})$ is the Green-Lagrangian strain tensor. The fiber contractile Cauchy stress in Eq. (5) can be transformed to the 2$^{nd}$ PK stress $\tilde{S}_{ij}^{FB}$ by dividing the stress from Eq. (5) by the stretch ratio in the fiber direction, and then rotate it to the Cartesian coordinate in the undeformed configuration.

Next, the viscous behavior is introduced to the deviatoric and volumetric components of muscle matrix using the continuous generalized Maxwell formulation [38] (see Appendix A for more details), and the final expressions are as follows.

$$S_{ij}^{dev}(t_{n+1}) \approx S_{ij,n+1}^{dev} = \tilde{S}_{ij,n+1}^{dev} + \sum_{p=1}^{N} \bar{g}_p H_{p,ij}^{n+1} \tag{11}$$

$$S_{ij,n+1}^{vol} = \tilde{S}_{ij,n+1}^{vol} + \sum_{p=1}^{N} \bar{b}_p L_{p,ij}^{n+1} \tag{12}$$

$$H_{p,ij}^{n+1} = \exp\left(-\frac{\Delta t}{\tau_p}\right) H_{p,ij}^n + \left(\frac{1-\exp\left(-\frac{\Delta t}{\tau_p}\right)}{\frac{\Delta t}{\tau_p}}\right) \left(\tilde{S}_{n+1,ij}^{dev} - \tilde{S}_{n,ij}^{dev}\right) \tag{13}$$

$$L_{p,ij}^{n+1} = \exp\left(-\frac{\Delta t}{\tau_p}\right) L_{p,ij}^n + \left(\frac{1-\exp\left(-\frac{\Delta t}{\tau_p}\right)}{\frac{\Delta t}{\tau_p}}\right) \left(\tilde{S}_{n+1,ij}^{vol} - \tilde{S}_{n,ij}^{vol}\right). \tag{14}$$

where $\bar{g}_p$ and $\bar{b}_p$ ($p = 1 \dots N$) are the *N*-term Prony-series deviatoric and volumetric relaxation coefficients, respectively, given in Table 3, and the derivation of $H_{p,ij}^{n+1}$ and $L_{p,ij}^{n+1}$ are given in Appendix A. Both deviatoric and volumetric Prony series use the same coefficients in our model.

Finally, the total 2$^{nd}$ PK stress at time step $n+1$ is then written as

$$S_{n+1,ij} = S_{n+1,ij}^{dev} + S_{n+1,ij}^{vol} + \tilde{S}_{n+1,ij}^{FB}. \tag{15}$$

The tangent obtained from the linearization process also needs to be updated at each time step.

$$C_{ijkl}^{n+1} = \frac{\partial S_{n+1,ij}}{\partial E_{kl}^{n+1}} = \frac{\partial S_{n+1,ij}^{dev}}{\partial E_{kl}^{n+1}} + \frac{\partial S_{n+1,ij}^{vol}}{\partial E_{kl}^{n+1}} + \frac{\partial \tilde{S}_{n+1,ij}^{FB}}{\partial E_{kl}^{n+1}} \tag{16}$$

The expressions of the tangent matrix components are given in Appendix B. Using stress and constitutive tensor updates, one can perform the standard nonlinear analysis.

Table 3: Deviatoric (and volumetric) relaxation and time coefficients from the five term Prony series for muscle tissue used in [27].

| Relaxation Coefficients | Value | Time Coefficients | Value (sec) |
|---|---|---|---|
| $\bar{g}_1(=\bar{b}_1)$ | 3.78 | $\tau_1$ | 0.6 |
| $\bar{g}_2(=\bar{b}_2)$ | 1.63 | $\tau_2$ | 6 |
| $\bar{g}_3(=\bar{b}_3)$ | 0.46 | $\tau_3$ | 30 |
| $\bar{g}_4(=\bar{b}_4)$ | 0.54 | $\tau_4$ | 60 |
| $\bar{g}_5(=\bar{b}_5)$ | 0.72 | $\tau_5$ | 300 |

*2.1.3 Correlation coefficient calculations*

We are interested in observing the apparent relations between the skeletal muscle force output, pressure, and the maximum principal, volumetric, and maximum shear strains at various activation levels. A way to quantify these is to perform a correlation analysis to investigate the relationships between these variables. Since the studied variables may be correlated non-monotonically, the Spearman's rank correlation coefficient is used to measure the extent of monotonic correlation between a pair of random variables $X$ and $Y$, defined as

$$r_s(X,Y) = \frac{cov(R(X), R(Y))}{\sqrt{var(R(X))}\sqrt{var(R(Y))}}. \qquad (17)$$

The variables $X$ and $Y$ are first converted to their ranked variables $R(x_i)$, $R(y_i)$ such that they are ranked according to the magnitude of the $i^{th}$ samples, $x_i \in X$ and $y_i \in Y$. The correlations are then measured between the ranked variables $R(X)$ and $R(Y)$. The Spearman's correlation coefficient ranges from -1 to +1, where $r_s = \pm 1$ indicates a perfectly monotonically increasing/decreasing relationship between $X$ and $Y$. For a monotonically increasing relationship ($r_s = 1$), as $X$ increases $Y$ also increases; however, the overall relationship between $X$ and $Y$ could be linear or higher order. Likewise, in a monotonically decreasing correlation ($r_s = -1$), as one variable increases the other decreases.

## 2.2 Numerical Model Setup

The continuum-scale model under isometric contractions is simulated using the software Abaqus (Dassault Systèmes, 2022). A generic muscle model with initial (i.e., prior to fiber activation) pennation angle $\theta = 47°$ and a thickness of 0.4 cm is shown in Figure 1(a) [23]. The top and bottom ends of the tendon are kept fixed to simulate isometric muscle contractions at different activation levels. The muscle components are modeled by a hybrid finite element formulation with a bilinear displacement field and a constant pressure field. The muscle simulations are quasi-static under the plane strain condition, which is consistent with the experimental observations that the out-of-plane deformation is relatively small compared to the in-plane deformation [21,23]. The muscle belly undergoes fifteen different activations, which are described in Table 4 and shown in Figure 1(c)-(e).

For the correlation metric discussed in Section 2.1.3, the quantities under investigation are extracted from five locations $(A, B, C, D, E)$ on the muscle belly as shown in Figure 1(a). These regions of interest (RoI) relate to muscle physiology; $A$, $B$ and $C$ are further away from the muscle belly, $D$ is at the middle of the belly, while $E$ is in the belly but bordering the aponeurosis. The force output is calculated from the reaction of the model at the bottom support. The pressure in the muscle belly during isometric contraction is defined as

$$p = -\frac{\partial W_{\text{MT}}^{\text{vol}}}{\partial J}. \tag{18}$$

The volumetric strain $\varepsilon_{vol}$ computed in the muscle belly is defined as

$$\varepsilon_{vol} = \text{tr}(\boldsymbol{\varepsilon}^N) \tag{19}$$

where $\boldsymbol{\varepsilon}^N = \sqrt{\mathbf{F} \cdot \mathbf{F}^T} - \mathbf{I}$ is the nominal strain. The principal strains are calculated from the nominal strains, where the maximum and minimum principal strains are denoted as $\epsilon_1$ and $\epsilon_3$, respectively. The maximum shear strain is then obtained as

$$\gamma_{max} = \frac{\epsilon_1 - \epsilon_3}{2}. \tag{20}$$

Table 4: The parameters of fifteen activation profiles for isometric contractions. $\bar{t}$ is the normalized time from 0 to 1.

| Profile Type | Equation | Parameters | |
|---|---|---|---|
| Linear | $a(\bar{t}; A_0) = A_0 \bar{t}$ | | $A_0 = 0.1$ |
| | | | $A_0 = 0.3$ |
| | | | $A_0 = 0.5$ |
| | | | $A_0 = 0.7$ |
| | | | $A_0 = 1.0$ |
| Non-linear | $a(\bar{t}; k, A_0) = A_0(1 - \exp(-k\bar{t}))$ | $k = 5$ | $A_0 = 0.1$ |
| | | | $A_0 = 0.3$ |
| | | | $A_0 = 0.5$ |
| | | | $A_0 = 0.7$ |
| | | | $A_0 = 1.0$ |
| | | $k = 10$ | $A_0 = 0.1$ |
| | | | $A_0 = 0.3$ |
| | | | $A_0 = 0.5$ |
| | | | $A_0 = 0.7$ |
| | | | $A_0 = 1.0$ |

# 3   Results and Discussions

## 3.1   *Force, pressure and strain outputs*

The total force from the muscle contraction is extracted from the reaction forces at the bottom support, and the outputs of pressure and strains are compared along the line segments $L_{long}$, $L_{short}$, $L_1$ and $L_2$, where the distances along those lines are measured starting from X=0 (see Figure 1 (b)).

<u>(i) Force Output</u>

The force output results shown in Figure 3 closely follow their respective activation profiles, as is evident from Figure 1 (c)-(e). For each activation level $A_0$, they reach similar levels of peak force output measured at the end of the simulation when compared between linear and non-linear ($k = 5, 10$) activation profiles. For the cases with non-linear activation, the rise to the peak force is faster than their linear counterparts, but they still arrive at the same peak force at the final deformation ($t = 1$).

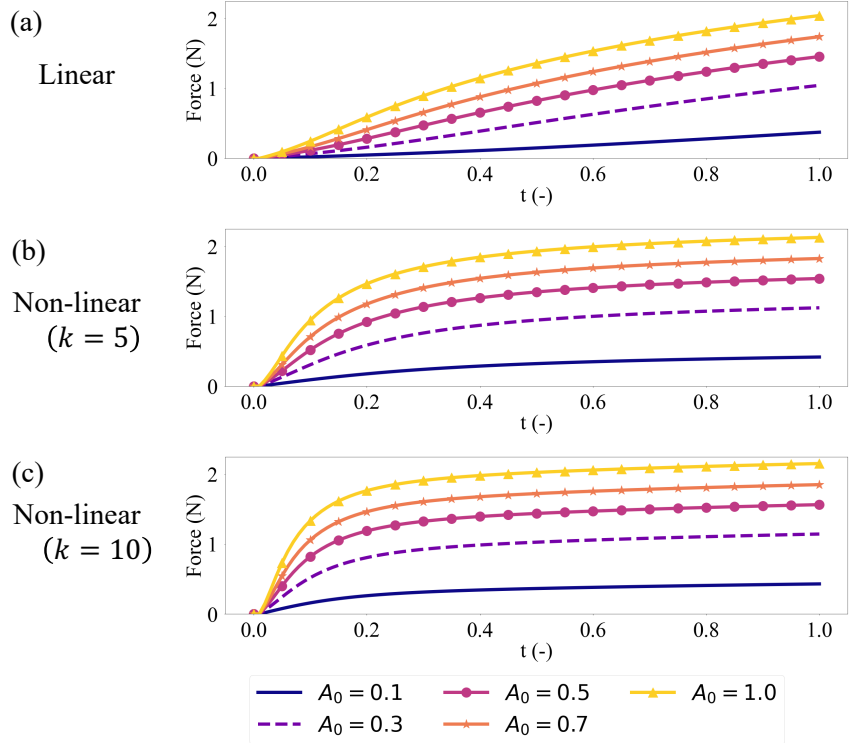

Figure 3: Force output from the skeletal muscle model vs simulation time ($t$) for the linear (a) and non-linear (b-c) activation profiles.

(ii) Pressure

For all cases along the long diagonal ($L_{long}$), the pressure increases and reaches the peak at the center of the belly as the time for activation increased from $t = 0.1$ to $1.0$, while the minimum pressure was observed at the interface of the aponeurosis and belly. As with force outputs in Figure 3, the pressure distributions at the final time-step ($t = 1.0$) are similar as we move from linear to non-linear cases for the same $A_0$, showing the path-independence of the material behavior. For non-linear activations (especially $k = 10$), the pressure distribution becomes stabilized and does not increase significantly beyond $t = 0.5$ for all activation magnitudes considered, due to the decreasing activation rate as time increases. A pressure relaxation is observed from $t = 0.5$ to $t = 1.0$ for $k = 5$ and $10$, where the pressure increase due to activation increment has been counterbalanced by pressure relaxation. Along the short diagonal ($L_{short}$), the pressure distributions are more uniform for all cases. As with the distributions along the long diagonal, the pressure stabilizes at $t = 0.5$ for the non-linear activation profiles due to pressure relaxation.

For distributions along the fiber direction near the traction-free surface close to the top of the muscle belly ($L_1$), low pressure is observed. Tensile pressure increases noticeably with time across all activation profiles near the aponeurosis-muscle belly interface. Again, pressure relaxation is observed for the non-linear activation cases. Similarly, for pressure distributions along the fiber direction near the middle of muscle belly ($L_2$), the pressure stabilizes for non-linear activations beyond $t = 0.5$ due to the counterbalancing effect of pressure relaxation.

The results for other sub-maximal activations ($A_0 = 0.3, 0.7$) are provided in Appendix C. Overall, these *in silico* results show that pressure has regional variation and requires careful selection of location to estimate the maximum IMP, which is an issue that has been observed by other *in vivo/ex vivo* studies [12,39]. The pressure relaxation effects lead to a weaker correlation between pressure and force output as will be discussed in the next section.

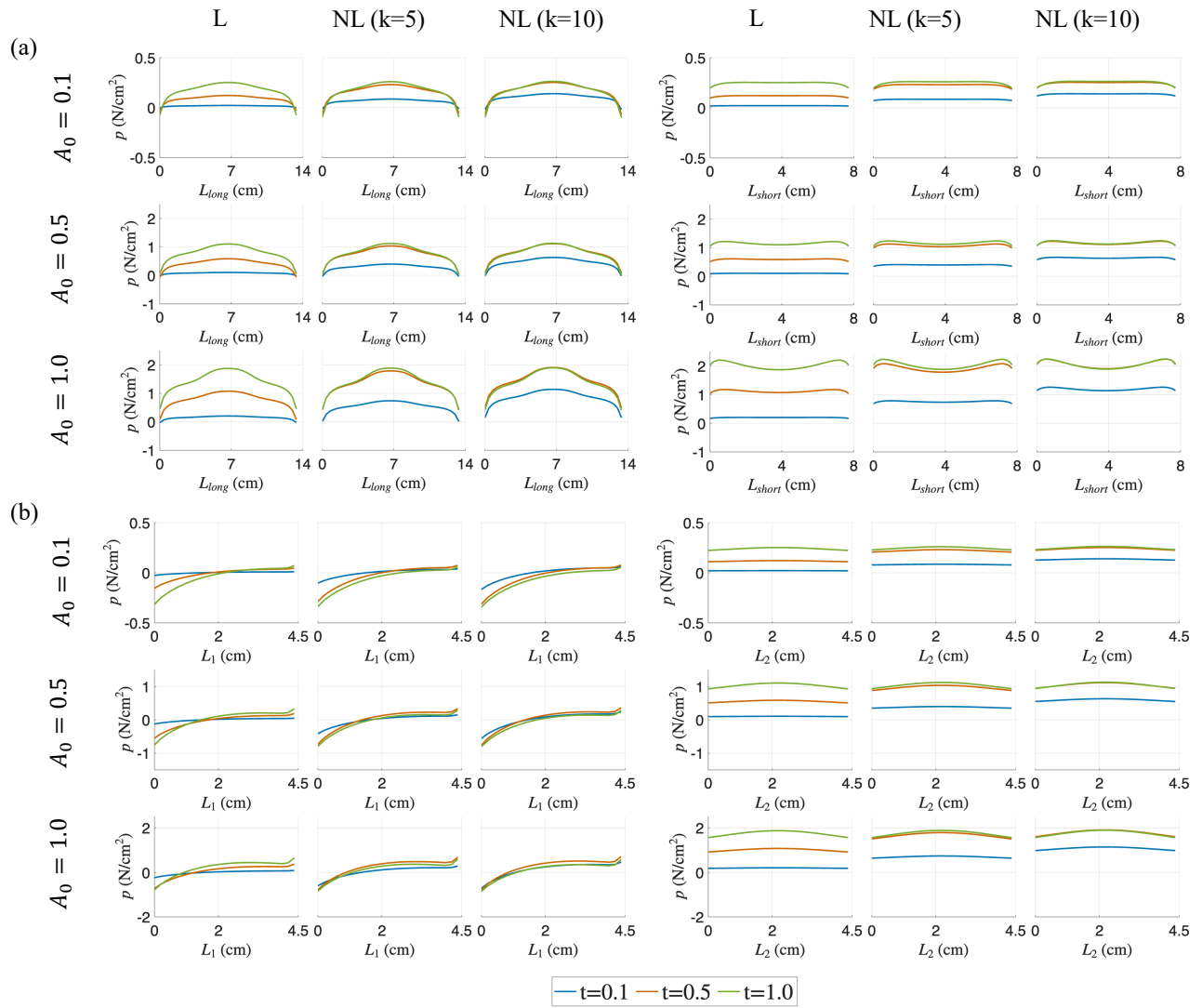

Figure 4: Pressure ($p$) distributions at three different simulation times ($t = 0.1, 0.5$ and $1.0$) for varying activation profiles (linear (L) and non-linear (NL)) and for $A_0 = 0.1, 0.5$ and $1.0$ in the muscle belly. The distributions along the diagonal lines are shown in (a) and the lines along the muscle fiber direction are shown in (b).

(iii) Volumetric, principal and maximum shear strains

As shown in Figure 5, the volumetric strain, $\varepsilon_{vol}$, increases throughout the belly along all chosen measurement lines (i.e., diagonals and along the fibers) as the activation increases with time, with a majority of the muscle belly undergoing an increase in the strain measure as the contraction progresses to its final state. Overall, we observe less spatial variations in volumetric strain compared to pressure distribution, except along the long diagonal $(L_{long})$ at a lower $A_0$. As with pressure and force observations, the strain distributions along each measurement line and for each $A_0$ case at the final time-step ($t = 1.0$) are similar between different activation profiles (linear, nonlinear with k = 5 and 10), demonstrating the path-independent material behavior. Unlike the pressure response, these volumetric strain distributions do not stabilize for all cases. The distributions of maximum principal stress $\epsilon_1$ and maximum shear strain $\gamma_{max}$ also show similar patterns (see Figure 6 and Figure 7). The results for other sub-maximal activations ($A_0 = 0.3, 0.7$) are provided in Appendix C.

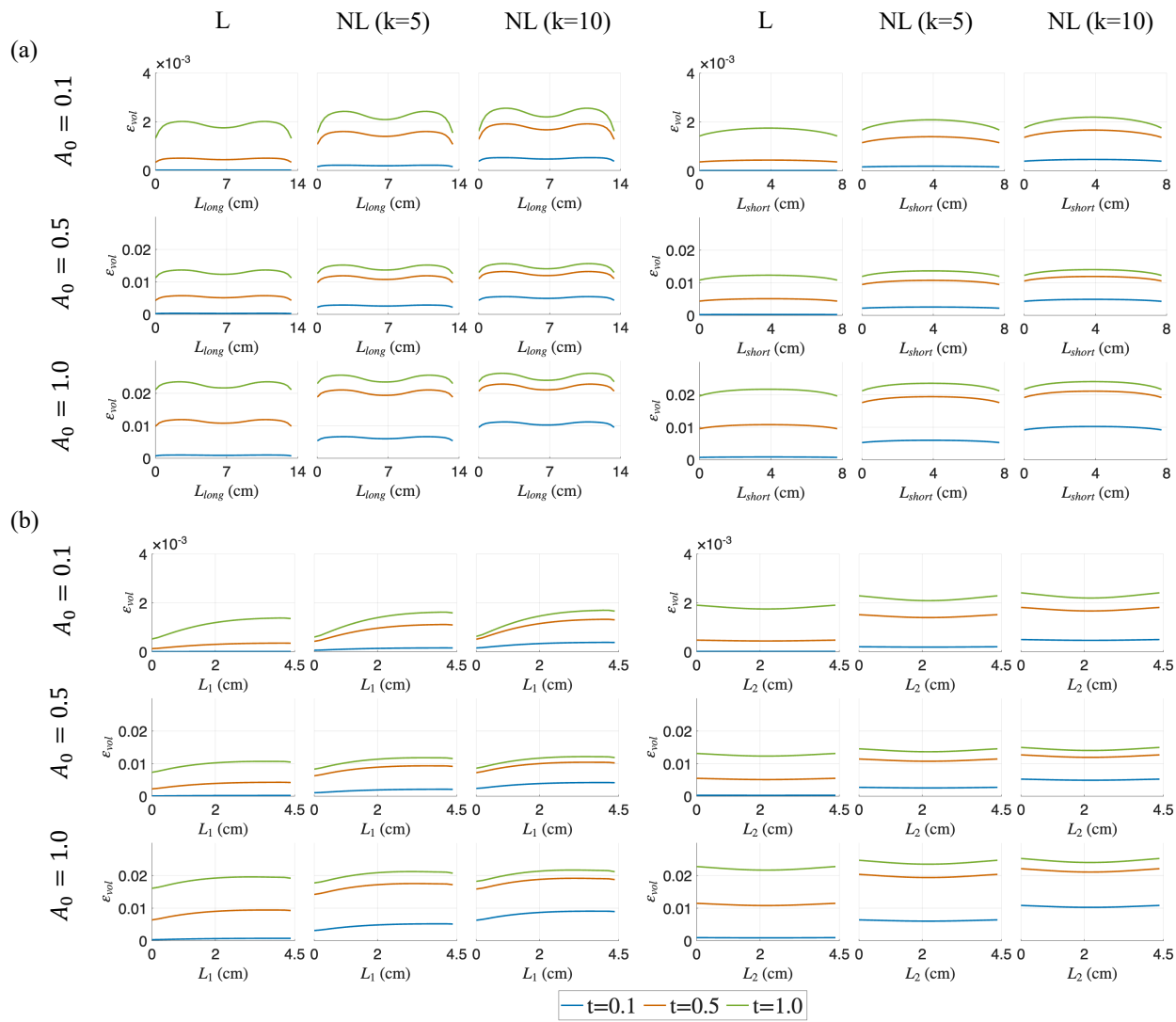

Figure 5: Volumetric strain ($\varepsilon_{vol}$) distributions at three different simulation times ($t = 0.1, 0.5$ and $1.0$) for varying activation profiles (linear (L) and non-linear (NL)), and for $A_0 = 0.1, 0.5$ and $1.0$ in the muscle belly. The distributions along the diagonal lines are shown in (a) and the lines along the muscle fiber direction are shown in (b).

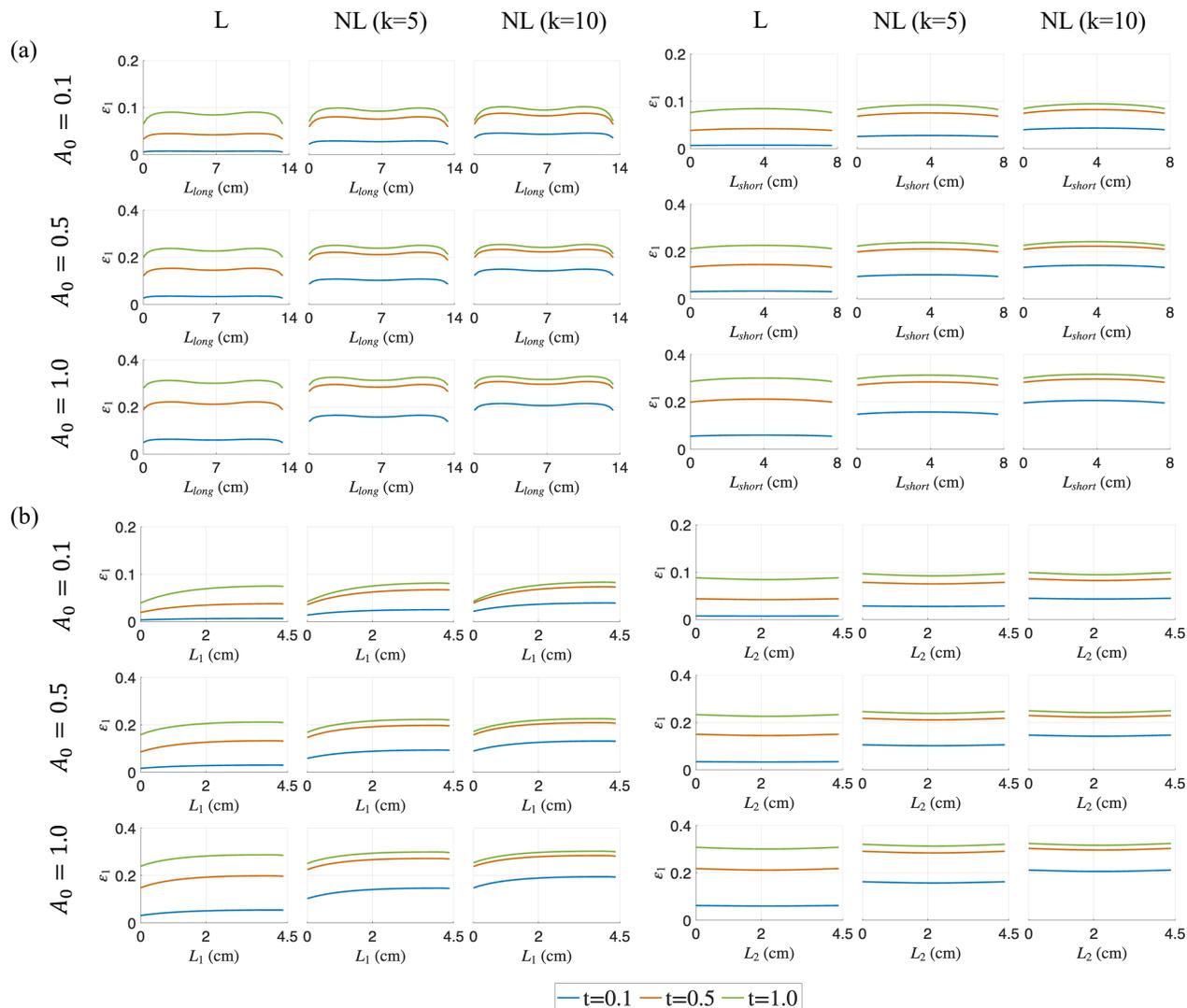

Figure 6: Maximum principal strain ($\varepsilon_1$) distributions at three different simulation times ($t = 0.1, 0.5$ and $1.0$) for varying activation profiles (linear (L) and non-linear (NL)), and for $A_0 = 0.1, 0.5$ and $1.0$ in the muscle belly. The distributions along the diagonal lines are shown in (a) and the lines along the muscle fiber direction are shown in (b).

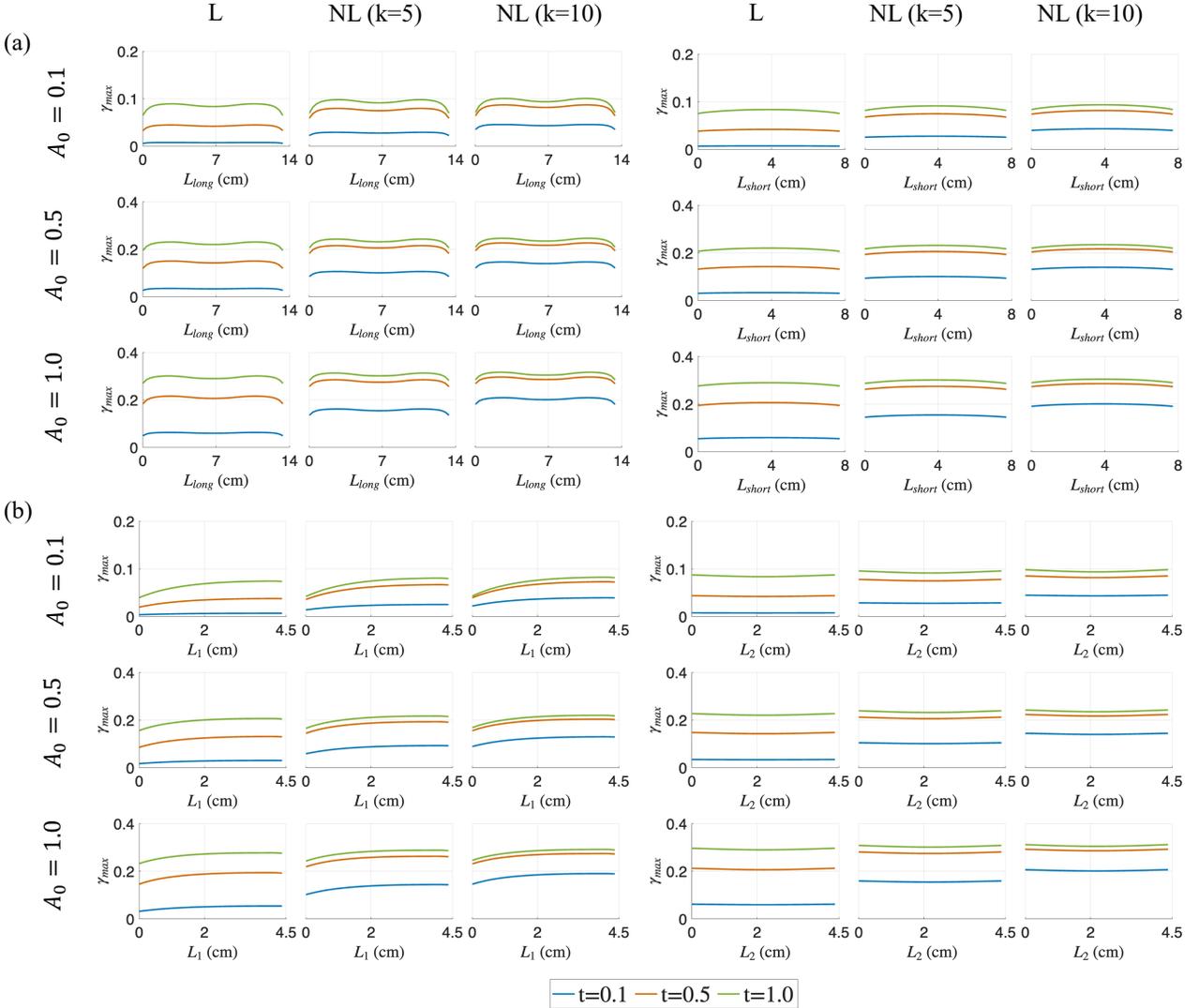

Figure 7: Maximum shear strain ($\gamma_{max}$) distributions at three different simulation times ($t = 0.1, 0.5$ and $1.0$) for varying activation profiles (linear (L) and non-linear (NL)), and for $A_0 = 0.1, 0.5$ and $1.0$ in the muscle belly. The distributions along the diagonal lines are shown in (a) and the lines along the muscle fiber direction are shown in (b).

## 3.2 Correlations between Force Output, Pressure, and Strain Measures

To further analyze the relationship between the variables of interest, the evolution of the force output, pressure, and strains (volumetric, maximum principal and maximum shear), at various activation levels ($A_0 = 0.1, 0.5$, and $1.0$) and profiles (linear and non-linear) were plotted pairwise with respect to each other in Figure 8(a), Figure 9(a), and Figure 10 (a). Their respective correlation coefficients, calculated as described in Section 2.1.3, are provided in Figure 8(b), Figure 9(b), and Figure 10(b).

A positive correlation indicates that as the independent variable increases, so does the dependent variable. For linear activation profiles, we observe strong positive correlations between force output and the various strain measures at all locations ($A \rightarrow E$). This trend is exhibited for non-linear activations also, where the volumetric, maximum principal and maximum shear strains versus force show a similar positive correlation for all activation levels $A_0$.

The pressure vs. volumetric strain and force vs. pressure relations, however, show weaker correlations at RoI's $(A, B, C)$ away from the muscle belly center for sub-maximal activations ($A_0 = 0.1 \rightarrow 0.7$). For $A_0 = 0.1$ and $0.3$, it is observed that these quantities at locations D and E show stronger monotonicity; as the strains increase, so does the pressure and the force output. This is not the case, however, at locations *A, B* and *C* which are located away from the center of the belly, with pressure as dependent or independent variable in Figure 8(a) – Figure 10(a), near the traction-free surface. These results may be explained by the occurrence of pressure stabilization/relaxation at these locations, as the force and strains reach their peaks. For non-linear activation profiles, the pressure stabilizes throughout the belly and relaxes in areas close to the aponeurosis interface near the ending stage of activation. The higher the initial activation rate, the more time there is for the pressure to stabilize due to relaxation near the traction-free surface. Moreover, as we move from $A_0 = 0.1$ (sub-maximal) to $A_0 = 1.0$ (maximal) activation levels, the stabilization-relaxation effect (see Figure 4) starts to affect the regions not only along the traction-free surface (*A, B* and *C*), but also in the interior regions of the belly (*D* and *E*). As such, weaker correlations through all the RoI's in the belly for $A_0 = 0.5 \rightarrow 1.0$ are observed.

Notably, in the case of linear activations where activation rate remains constant, the pressure relaxation effect is absent for all peak activation magnitudes ($A_0$). In fact, we notice stronger correlations for linear cases as we move from sub-maximal to maximal activations. This suggests that besides the level of activation, activation rate is also an important factor when investigating force output. From visco-hyperelastic theory, slower deformations lead to responses closer to quasi-hyperelasticity with minimal stress-relaxation. This agrees with what is observed from our simulations.

These effects are further quantified by calculated correlation coefficients in sub-panels (b) of Figure 8–Figure 10. For non-linear activation profiles, we observe $r_s = +1$ between the deformation measures (i.e., volumetric, max principal and max shear strains) and force for all simulated cases. Due to the pressure stabilization-relaxation effects, we see negative or zero $r_s$ for correlations with pressure for smaller sub-maximal activations at locations *A*, *B* and *C*, while *D* and *E* show strong positive $r_s$. As activation levels increase to maximum, these non-positive numbers become weaker, reaching smaller $r_s$ values (either positive or negative) at all locations. For linear activations, we see positive correlations for all simulated cases, with $r_s$ close to 1.0. The results for other sub-maximal activations ($A_0 = 0.3, 0.7$) are provided in Appendix D.

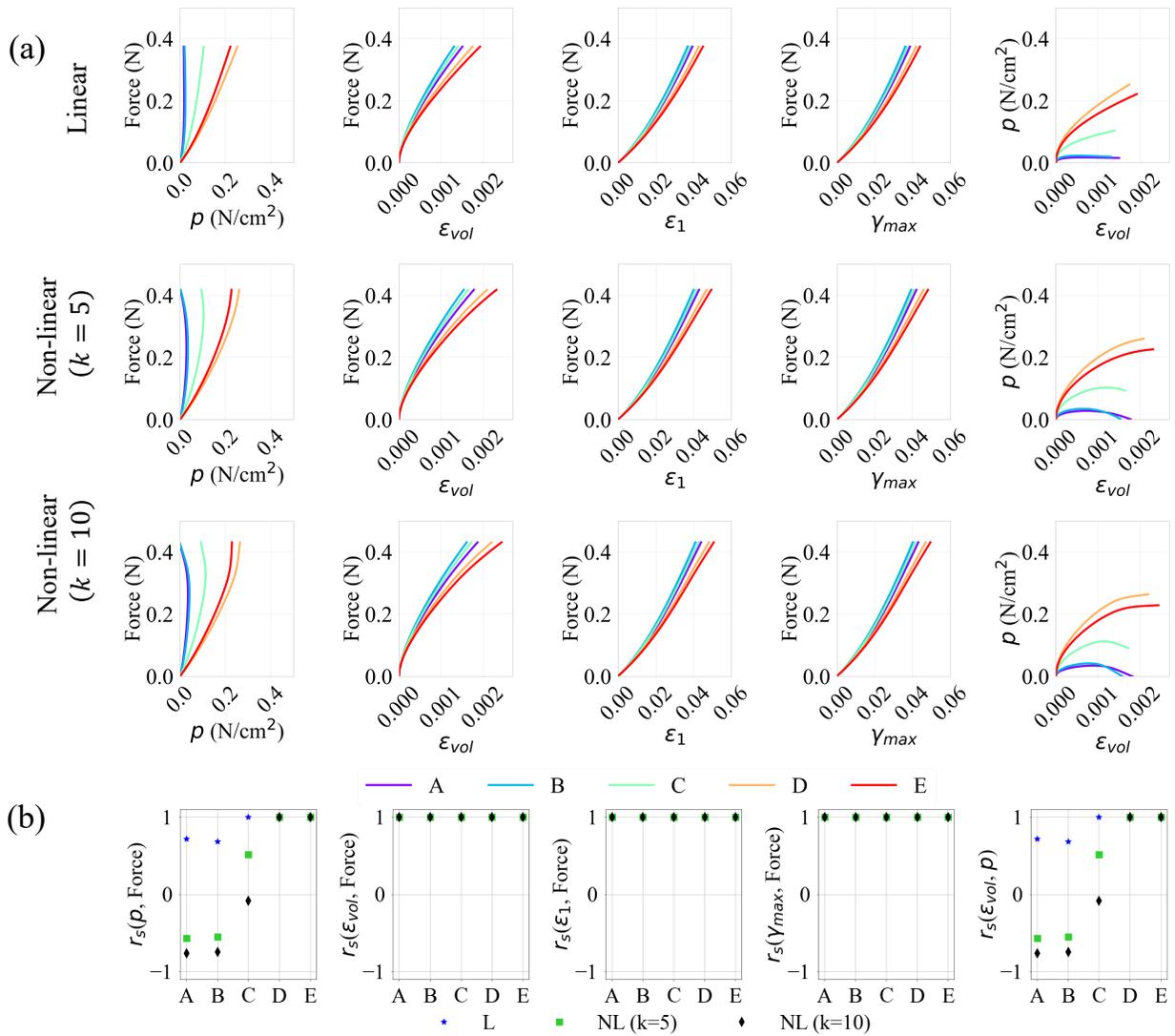

Figure 8: Evolution of relationships between the force output, pressure, and maximum principal, maximum shear and volumetric strains (a), and their spearman correlation coefficients (b), for linear (L) and non-linear (NL) activation profiles, for a maximal activation of "$A_0 = 0.1$", at locations $A - E$ on the muscle belly.

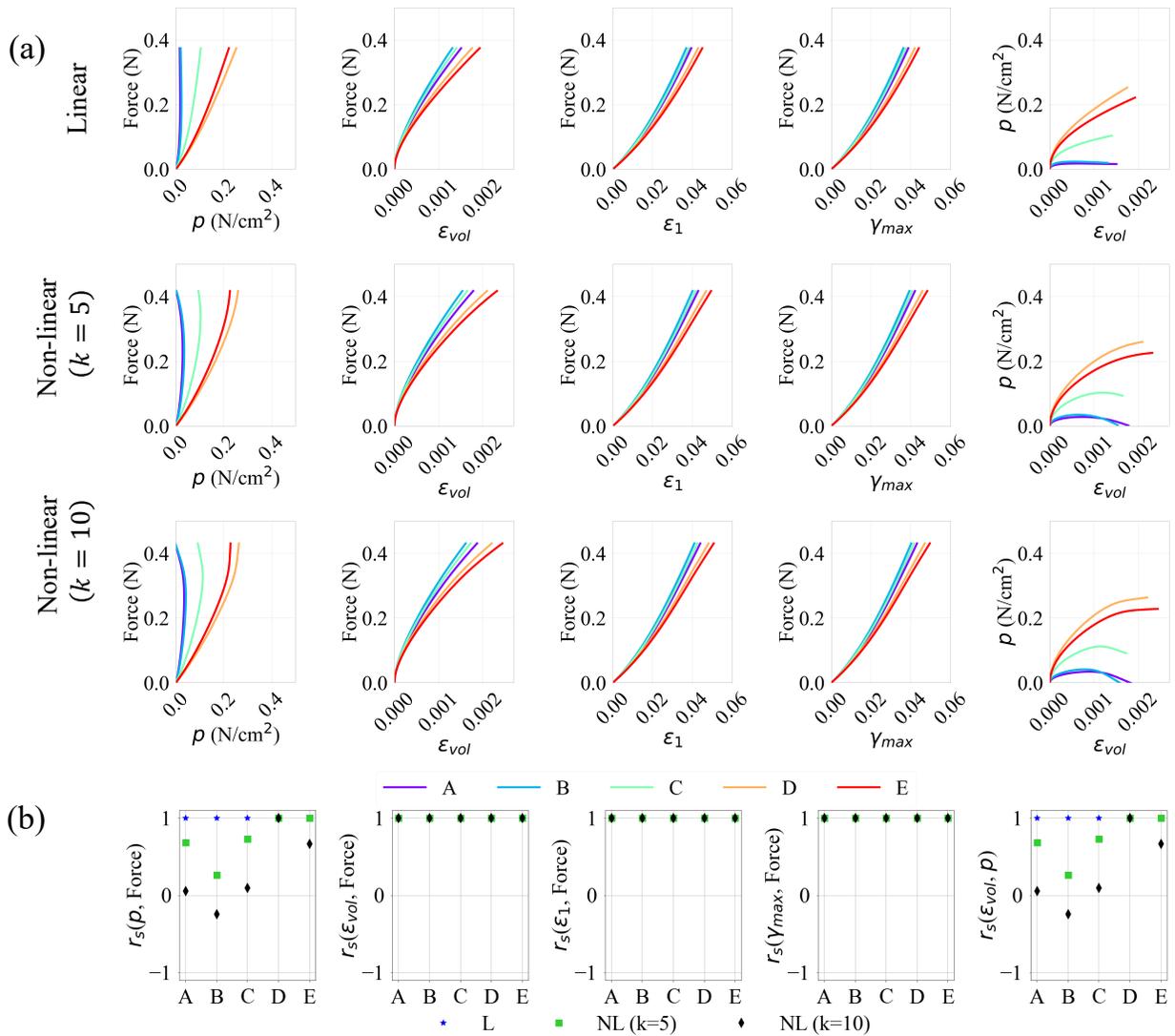

Figure 9: Evolution of relationships between the force output, pressure, and maximum principal, maximum shear and volumetric strains (a), and their spearman correlation coefficients (b), for linear (L) and non-linear (NL) activation profiles, for a maximal activation of "$A_0 = 0.5$", at locations $A - E$ on the muscle belly.

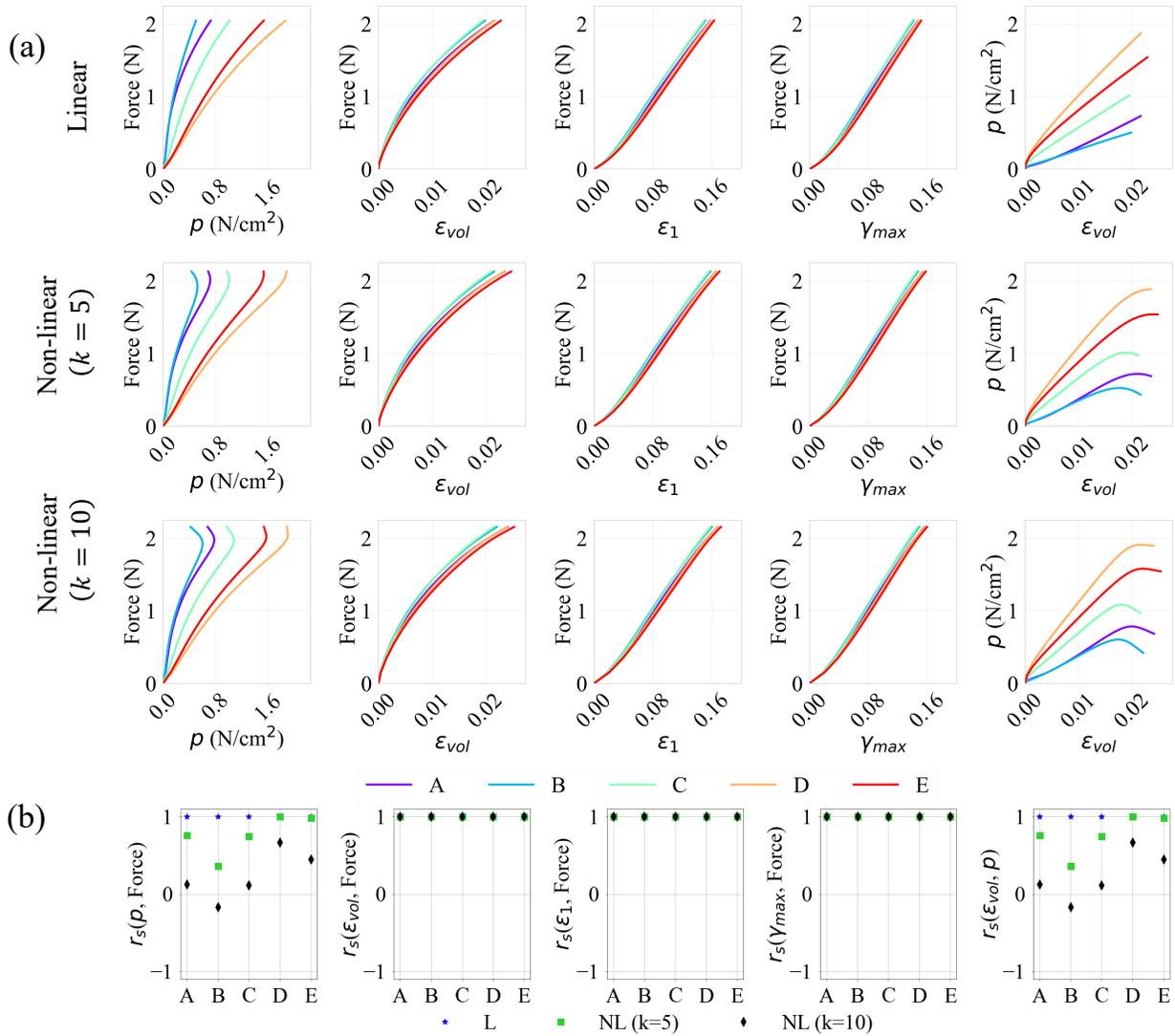

Figure 10: Evolution of relationships between the force output, pressure, and maximum principal, maximum shear and volumetric strains (a), and their spearman correlation coefficients (b), for linear (L) and non-linear (NL) activation profiles, for a maximal activation of "$A_0 = 1.0$", at locations $A - E$ on the muscle belly.

In summary, these results indicate that as force increases during muscle contraction, the maximum principal, maximum shear and volumetric strain increase at all locations, indicated by the strong Spearman's correlation coefficients as shown Figure 8(b)–Figure 10(b). On the contrary, force-pressure and pressure-volumetric strain do not correlate well at the traction-free surface and at the locations near the boundary between the muscle belly and aponeurosis due to pressure stabilization and relaxation. It is, however, evident that the strongest correlation between force, pressure, and strains are at the center of the muscle belly, indicating a potentially best location to correlate these quantities for sub-maximal activations. For maximal activations, the correlations between force and strains are strong, but weaker for force and pressure. These results, which are consistent with those observed in other experimental and computational studies [13,21,31], have a major implication on the experimental design and measurement location for strains and force measurements.

## 4 Discussion and Conclusion

Surrogate measures for contributions of individual muscle forces in a muscle group have traditionally been measured through the electrical activity of the muscle (via electromyograms (EMG)) and intramuscular pressure (IMP) [5,27]. However, the relationship between muscle force and EMG/IMP is often non-linear, reflecting the dynamic changes in muscle recruitment throughout movement. This necessitates mechanics-based frameworks to account for varying contributions of multiple muscles to joint torque over time [4,9]. Furthermore, EMG/IMP experienced by the muscle belly during contractions can only be monitored by sensors that are embedded invasively (albeit minimal). To explore the alternative and more reliable proxies for muscle force, we used simulations to investigate the correlation between various strain measures, pressure, and force output, with strain measures as potential alternatives that are non-invasive and easier to apply in the MRI environment [13,14].

In this study, a visco-hyperelastic model has been used in the skeletal muscle modeling to investigate the correlations between the force output, various strain measures (volumetric, maximum principal and maximum shear strains), and the pressure of the continuum skeletal muscle. This numerical simulation followed by correlation analysis on these quantities provides insight into the dependence of pressure and strains on the force output. The numerical investigation of such a relationship can be used to establish strain metrics that have better correlations to muscle force output for future experimental studies.

The skeletal muscle model is subjected to isometric contraction with a range of linear and non-linear activation profiles with sub-maximal and maximal activation magnitudes. The distribution patterns of pressure and volumetric strain appear qualitatively similar (Figure 4 and Figure 5). It is also clear from these figures and from the separation of the regional plots in Figure 8(a)–Figure 10(a) that the variation of strain metrics across the muscle is less than 20% whereas pressure varies by more than 100% across the muscle, suggesting that strains may be less susceptible to changes in sample location.

Spearman's correlation coefficient was calculated to investigate the correlations between pressure, various strain measures, and the force output. It is observed that the strongest correlations between these variables happen at the center of the belly for both linear and non-linear activations, and for sub-maximal activations ($A_0 = 0.1, 0.3$ and $0.5$). However, pressure correlates well with other measures only for lower sub-maximal ($A_0 = 0.1$ and $0.3$) activations with shorter contraction period at peak (linear, and non-linear with $k = 5$). This is due to the stronger pressure stabilization-relaxation effect for higher sub-maximal activations that have higher initial activation rate (non-linear with $k = 10$). On the other hand, a strong correlation exists between all the strain measures (maximum principal, shear and volumetric) and force output at all locations of the belly, irrespective of the activation profiles. Since strains are non-invasively measurable, e.g., through MRI, this observation provides a pathway for better estimating the force in an individual muscle experimentally.

In future studies, 3-D continuum-scale muscle models constructed from realistic geometries will be incorporated. We will also consider both active and passive contractions as well as muscle shortening/lengthening to obtain a complete correlation of the contractile behaviors of the fiber with the force output.

## 5 Conflict of Interest Statement

The authors of this paper have no financial or personal relationships with other people or organizations that could inappropriately influence (bias) our work.

## 6 Funding Statement

The support of this work by the National Institute of Health under grant number 5R01AG056999 to the University of California San Diego is very much appreciated.

**Appendix A: Generalized Maxwell Visco-Hyperelastic Formulations**

The visco-hyperelastic stress formulations are derived from the generalized Maxwell model [24]. The deviatoric stresses are derived by the following equation, with superscript '$\sim$' denoting the hyperelastic contributions.

$$S_{ij}^{\text{dev}}(t) = \tilde{S}_{ij}^{\text{dev}}(t) + \sum_{p=1}^{N} \bar{g}_p \, H_{p,ij}(t) \tag{21}$$

The deviatoric visco-hyperelastic stress variables are written as

$$H_{p,ij}(t) = \int_0^t \exp\left[-\left(\frac{t-s}{\tau_p}\right)\right]\left(\frac{\partial \tilde{S}_{ij}^{\text{dev}}(s)}{\partial s}\right) ds \tag{22}$$

where $\bar{g}_p$ and $\tau_p$ $(p = 1, 2, \cdots, N)$ are the $N$-term Prony-series deviatoric coefficients and relaxation times. To get the stress update equation in the discrete form, we evaluate the expression at time $t_{n+1}$ by decomposing the integral form,

$$\begin{aligned}
H_{p,ij}(t_{n+1}) &= \int_0^{t_{n+1}} \exp\left[-\left(\frac{t_{n+1}-s}{\tau_p}\right)\right]\frac{d}{ds}\left(\tilde{S}_{ij}^{\text{dev}}(s)\right) ds \\
&= \int_0^{t_n} \exp\left[-\left(\frac{t_n-s}{\tau_p}\right)\right]\frac{d}{ds}\left(\tilde{S}_{ij}^{\text{dev}}(s)\right) ds \\
&\quad + \int_{t_n}^{t_{n+1}} \exp\left[-\left(\frac{t_{n+1}-s}{\tau_p}\right)\right]\frac{d}{ds}\left(\tilde{S}_{ij}^{\text{dev}}(s)\right) ds \\
&= \exp\left(-\frac{\Delta t}{\tau_p}\right) H_{p,ij}(t_n) \\
&\quad + \int_{t_n}^{t_{n+1}} \exp\left[-\left(\frac{t_{n+1}-s}{\tau_p}\right)\right]\frac{d}{ds}\left(\tilde{S}_{ij}^{\text{dev}}(s)\right) ds
\end{aligned} \tag{23}$$

where $\Delta t = t_{n+1} - t_n$. By using the first-order derivative approximation for the stress rate inside the integral [25],

$$\begin{aligned}
\int_{t_n}^{t_{n+1}} \exp\left[-\left(\frac{t_{n+1}-s}{\tau_p}\right)\right]\frac{d}{ds}\left(\tilde{S}_{ij}^{\text{dev}}(s)\right) ds \\
= \frac{S_{\text{dev},ij}^{(h,n+1)} - S_{\text{dev},ij}^{(h,n)}}{\Delta t}\left(\tau_p \exp\left[-\left(\frac{t_{n+1}-s}{\tau_p}\right)\right]\Big|_{s=t_n}^{s=t_{n+1}}\right) \\
= \left(\frac{1-\exp\left(-\frac{\Delta t}{\tau_p}\right)}{\frac{\Delta t}{\tau_p}}\right)\left(\tilde{S}_{n+1,ij}^{\text{dev}} - \tilde{S}_{n,ij}^{\text{dev}}\right),
\end{aligned} \tag{24}$$

we obtain a recursive expression for the visco-hyperelastic deviatoric internal stress variables.

$$H_{p,ij}(t_{n+1}) \approx H_{p,ij}^{n+1} = \exp\left(-\frac{\Delta t}{\tau_p}\right) H_{p,ij}^n + \left(\frac{1 - \exp\left(-\frac{\Delta t}{\tau_p}\right)}{\frac{\Delta t}{\tau_p}}\right) \left(\tilde{S}_{n+1,ij}^{\text{dev}} - \tilde{S}_{n,ij}^{\text{dev}}\right) \quad (25)$$

This leads to the discrete N-term Prony series stress update equation for the deviatoric stress tensor.

$$S_{ij}^{\text{dev}}(t_{n+1}) \approx S_{ij,n+1}^{\text{dev}} = \tilde{S}_{ij,n+1}^{\text{dev}} + \sum_{p=1}^{N} \bar{g}_p H_{p,ij}^{n+1} \quad (26)$$

Similarly, the visco-hyperelastic stress update equation for the volumetric stress can be derived,

$$S_{ij,n+1}^{\text{vol}} = \tilde{S}_{ij,n+1}^{\text{vol}} + \sum_{p=1}^{N} \bar{b}_p L_{p,ij}^{n+1} \quad (27)$$

where $\bar{b}_p$ ($p = 1 \to N$) are the *N*-term Prony-series volumetric relaxation coefficients and the volumetric internal stress variables are,

$$L_{p,ij}^{n+1} = \exp\left(-\frac{\Delta t}{\tau_p}\right) L_{p,ij}^n + \left(\frac{1 - \exp\left(-\frac{\Delta t}{\tau_p}\right)}{\frac{\Delta t}{\tau_p}}\right) \left(\tilde{S}_{n+1,ij}^{\text{vol}} - \tilde{S}_{n,ij}^{\text{vol}}\right). \quad (28)$$

**Appendix B: Visco-Hyperelastic Tangent Matrix Formulation**

The deviatoric tangent matrix is computed as

$$\begin{aligned}
\frac{\partial S_{ij,n+1}^{\text{dev}}}{\partial E_{kl,n+1}} &= \frac{\partial \tilde{S}_{ij,n+1}^{\text{dev}}}{\partial E_{kl,n+1}} + \sum_{p=1}^{N} \bar{g}_p \left(\frac{1 - \exp\left(-\frac{\Delta t}{\tau_p}\right)}{\frac{\Delta t}{\tau_p}}\right) \frac{\partial \tilde{S}_{ij,n+1}^{\text{dev}}}{\partial E_{kl,n+1}} \\
&= \left(1 + \sum_{p=1}^{N} \bar{g}_p \left(\frac{1 - \exp\left(-\frac{\Delta t}{\tau_p}\right)}{\frac{\Delta t}{\tau_p}}\right)\right) \frac{\partial \tilde{S}_{ij,n+1}^{\text{dev}}}{\partial E_{kl,n+1}}.
\end{aligned} \quad (29)$$

Similarly, the volumetric component update is written as,

$$\frac{\partial S^{\text{vol}}_{ij,n+1}}{\partial E_{kl,n+1}} = \left(1 + \sum_{p=1}^{N} \bar{b}_p \left(\frac{1 - \exp\left(-\frac{\Delta t}{\tau_p}\right)}{\frac{\Delta t}{\tau_p}}\right)\right) \frac{\partial \tilde{S}^{\text{vol}}_{ij,n+1}}{\partial E_{kl,n+1}}.$$

(30)

## Appendix C: Distributions from simulations for $A_0 = 0.3, 0.7$

The distributions of pressure $(p)$, volumetric $(\varepsilon_{vol})$, maximum principal $(\epsilon_1)$ and maximum shear $(\gamma_{max})$ strains are shown in Figure 11-Figure 14.

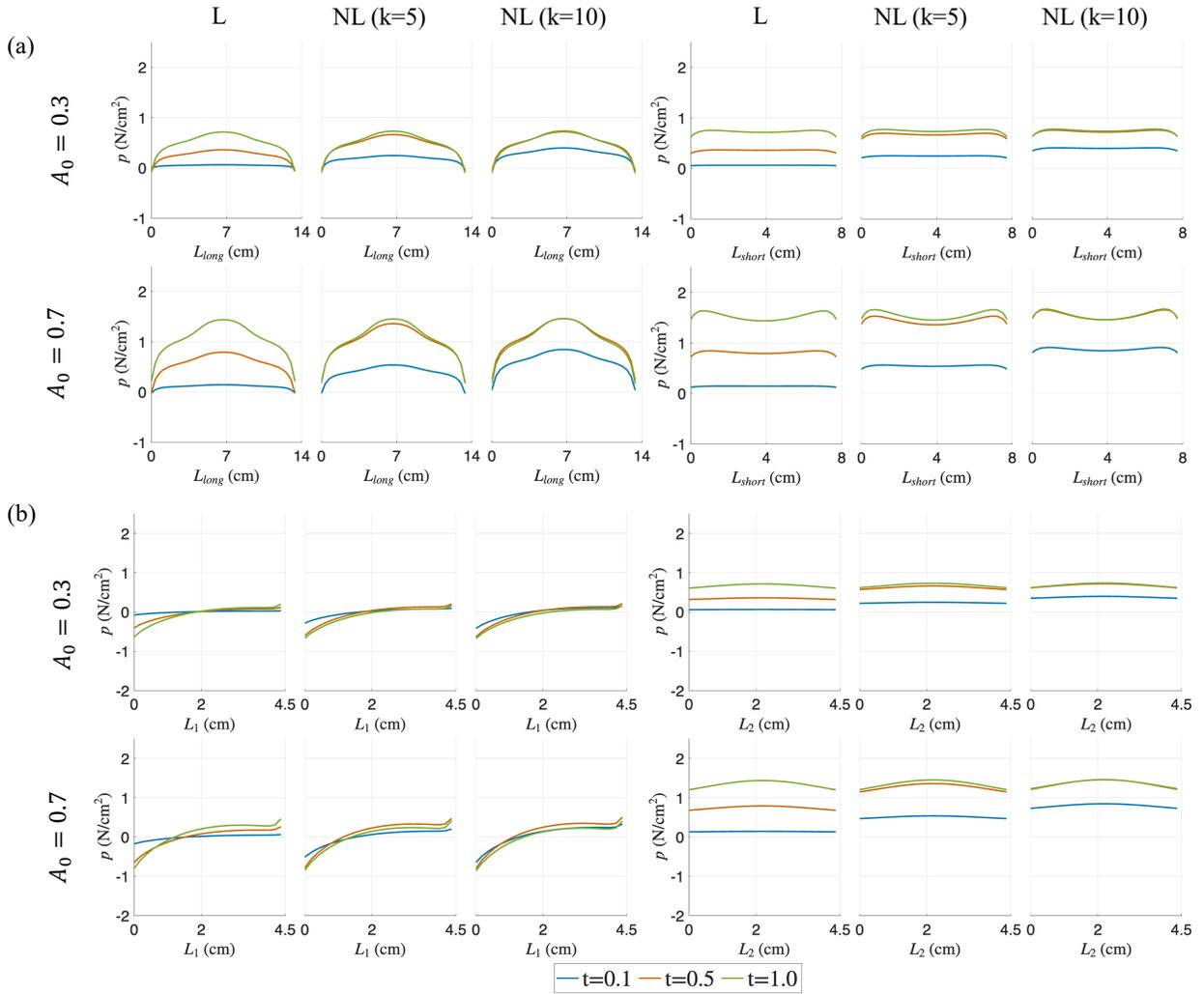

Figure 11: Pressure $(p)$ distributions at three different simulation times ($t = 0.1, 0.5$ and $1.0$) for varying activation profiles (linear (L) and non-linear (NL)) for $A_0 = 0.3$ and $0.7$ in the muscle

belly. The distributions along the diagonal lines are shown in (a) and the lines along the muscle fiber direction are shown in (b).

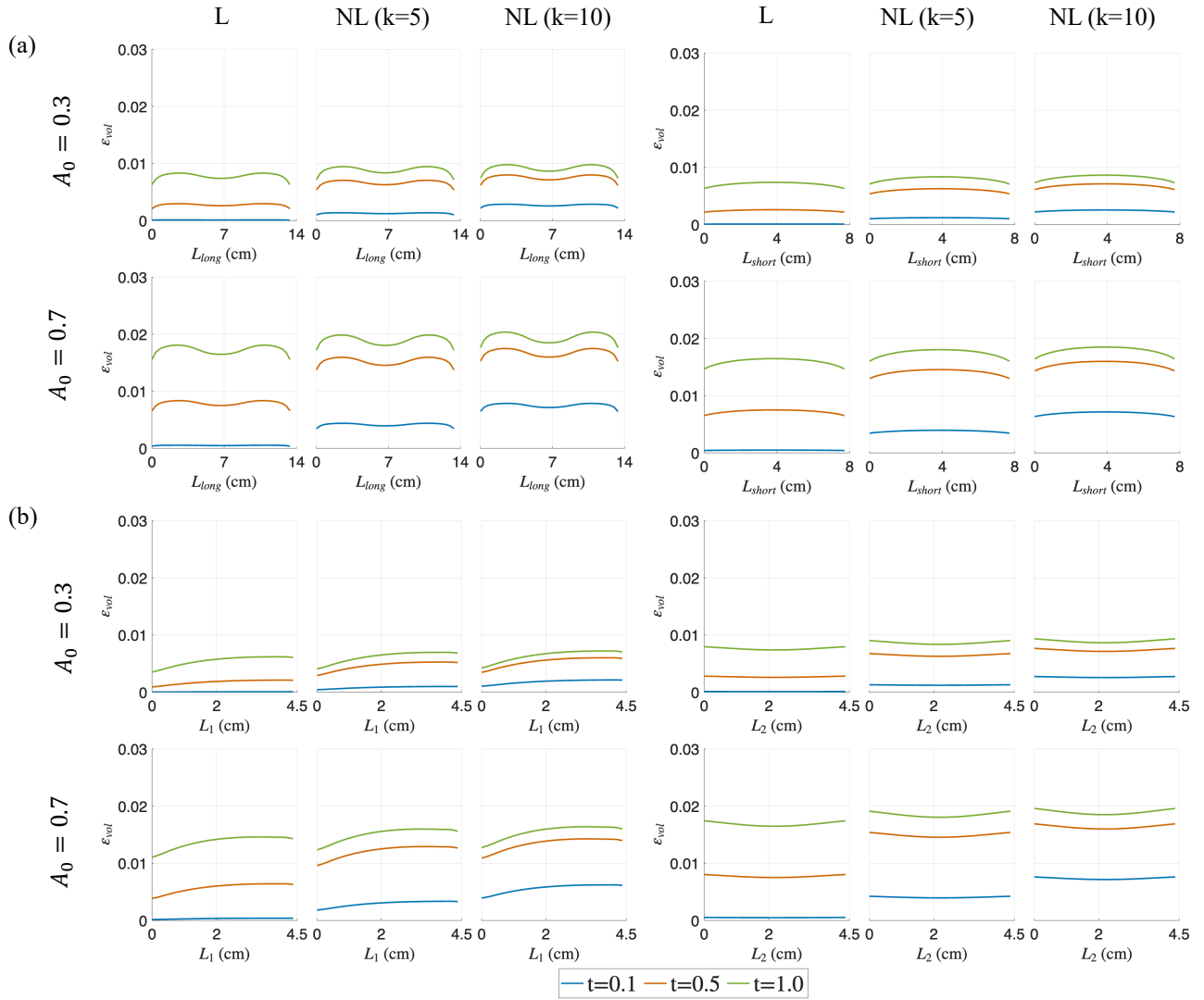

Figure 12: Volumetric strain ($\varepsilon_{vol}$) distributions at three different simulation times ($t = 0.1, 0.5$ and $1.0$) for varying activation profiles (linear (L) and non-linear (NL)) for $A_0 = 0.3$ and $0.7$ in the muscle belly. The distributions along the diagonal lines are shown in (a) and the lines along the muscle fiber direction are shown in (b).

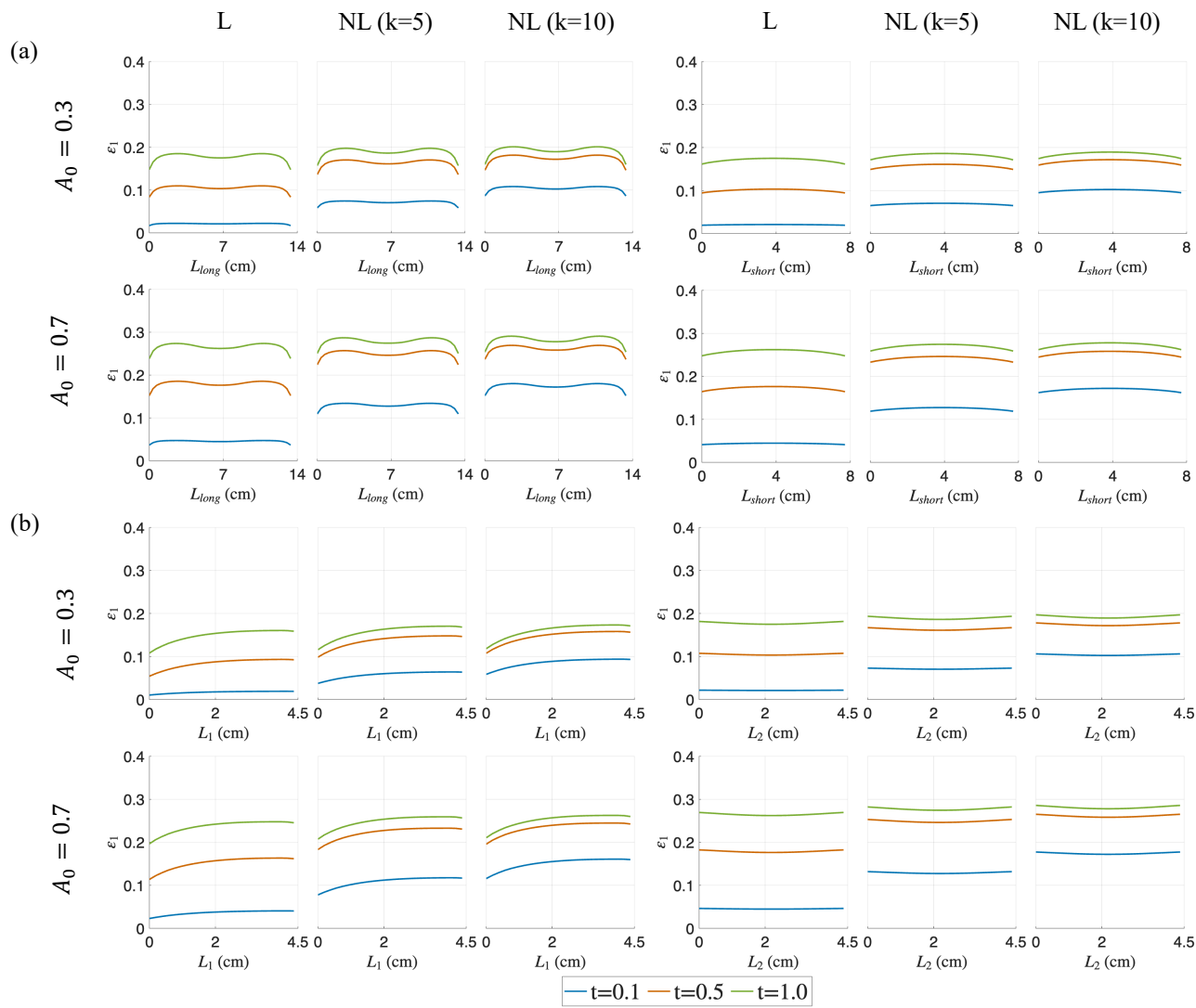

Figure 13: Maximum principal strain ($\varepsilon_1$) distributions at three different simulation times ($t = 0.1, 0.5$ and $1.0$) for varying activation profiles (linear (L) and non-linear (NL)) for $A_0 = 0.3$ and $0.7$ in the muscle belly. The distributions along the diagonal lines are shown in (a) and the lines along the muscle fiber direction are shown in (b).

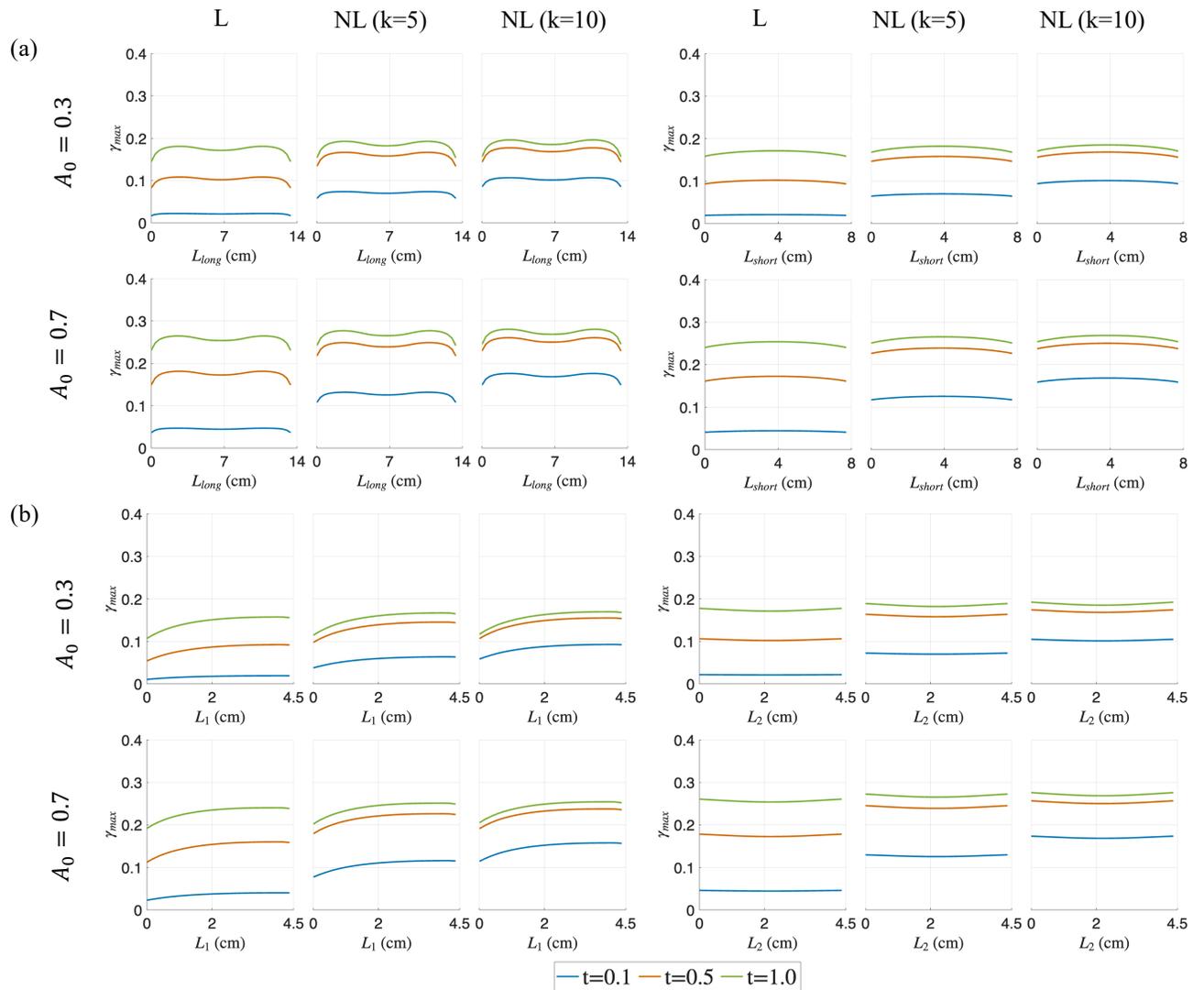

Figure 14: Maximum shear strain ($\gamma_{max}$) distributions at three different simulation times ($t = 0.1, 0.5$ and $1.0$) for varying activation profiles (linear (L) and non-linear (NL)) for $A_0 = 0.3$ and $0.7$ in the muscle belly. The distributions along the diagonal lines are shown in (a) and the lines along the muscle fiber direction are shown in (b).

**Appendix D: Correlation Plots for $A_0 = 0.3$ and $A_0 = 0.7$**

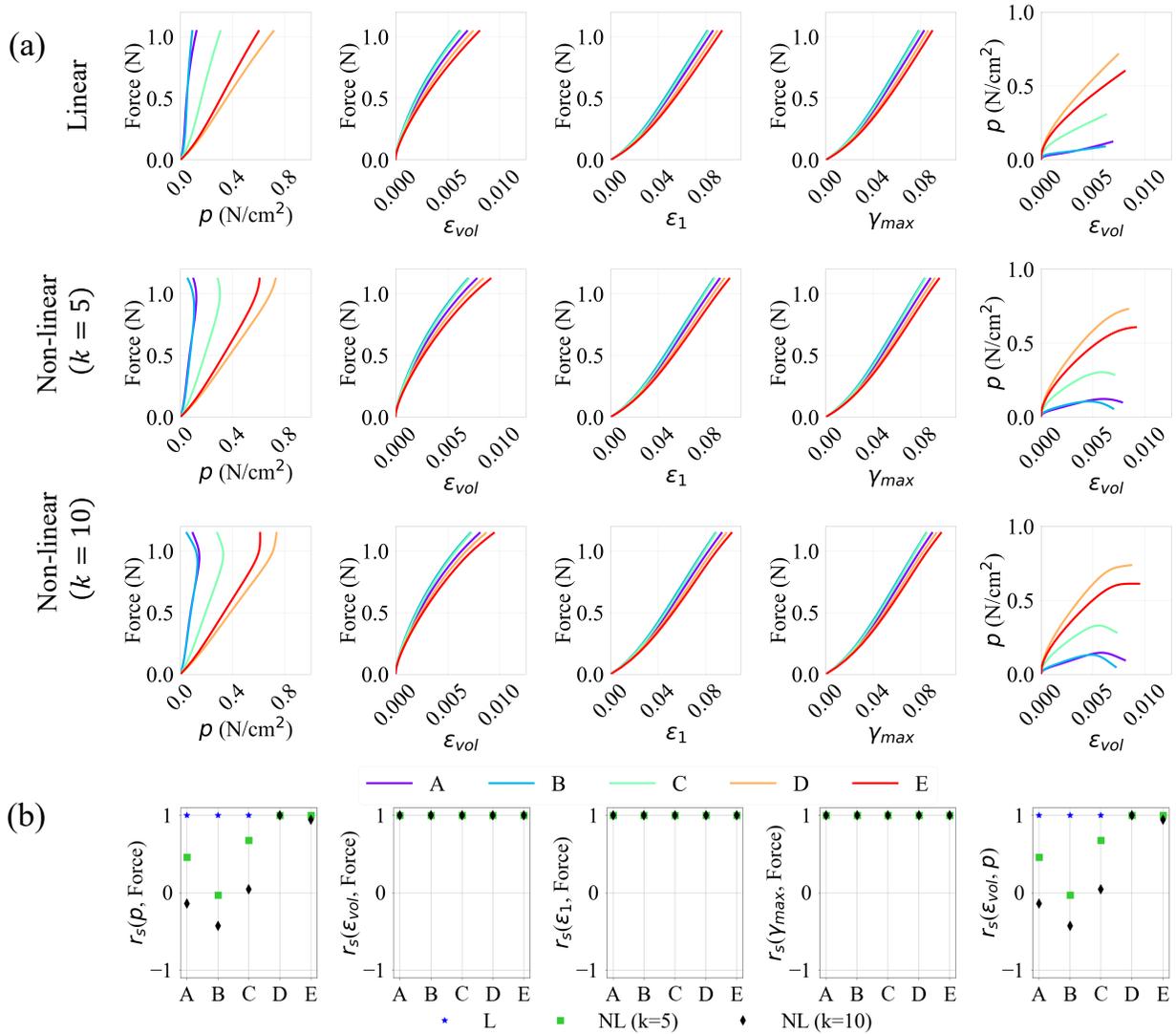

Figure 15: Evolution of the force output, pressure and maximum principal, maximum shear and volumetric strains (a), and their spearman correlation coefficients (b), for linear and non-linear activation profiles, for a maximal activation of "$A_0 = 0.3$", at locations $A - E$ on the muscle belly.

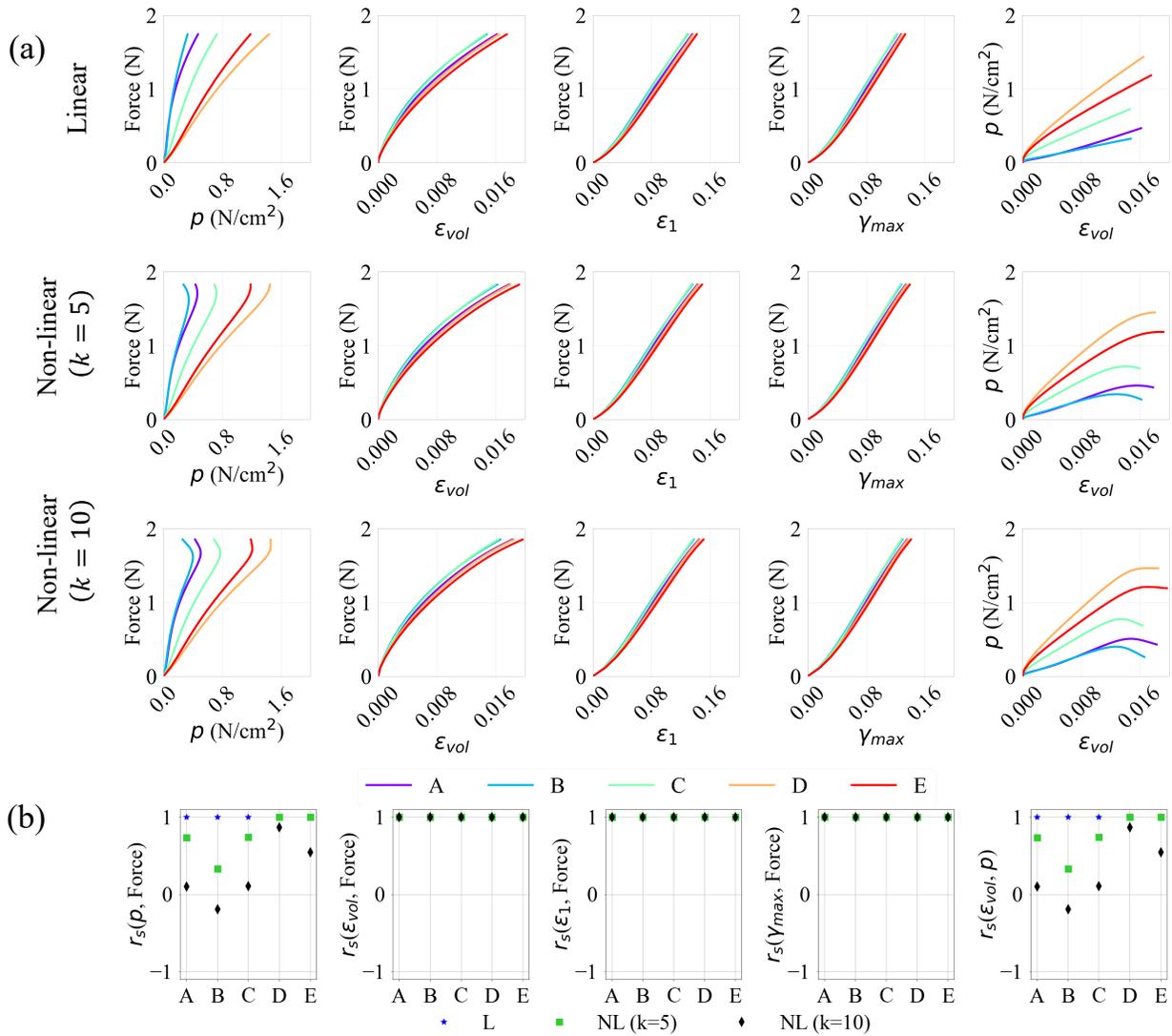

Figure 16: Evolution of the force output, pressure and maximum principal, maximum shear and volumetric strains (a), and their spearman correlation coefficients (b), for linear (L) and non-linear (NL) activation profiles, for a maximal activation of "$A_0 = 0.7$", at locations $A - E$ on the muscle belly.

**Figure Captions List**

Figure 1     An overview of the proposed modeling of the skeletal muscle. The model in (a) is a continuum-scale plane strain model with an initial pennation angle of 47° and a thickness of 0.4 cm in the Z-direction, where the dash-dot lines represent the orientation of muscle fibers in (b). The lines $L_1, L_2$ along the direction of the muscle fibers, and $L_{long}, L_{short}$ along the diagonals across the belly are the directions along which distributions of pressure and strains are examined. The five positions $A \rightarrow E$ in the belly indicate the locations where the pressure and strains are extracted for the statistical analysis. The plots in (c-e) show the fifteen different activation profiles described in Table 4.

Figure 2     Normalized (a) length-dependent and (b) velocity-dependent mechanical responses of the muscle fiber model with $a(\bar{t}) = 1$. The stretch and stretch rate at which the shortening and lengthening phases of the fiber exist are also indicated.

Figure 3   Force output from the skeletal muscle model vs simulation time ($t$) for the linear (a) and non-linear (b-c) activation profiles.

Figure 4   Pressure ($p$) distributions at three different simulation times ($t = 0.1, 0.5$ and $1.0$) for varying activation profiles (linear (L) and non-linear (NL)), and for $A_0 = 0.1, 0.5$ and $1.0$ in the muscle belly. The distributions along the diagonal lines are shown in (a) and the lines along the muscle fiber direction are shown in (b).

Figure 5   Volumetric strain ($\varepsilon_{vol}$) distributions at three different simulation times ($t = 0.1, 0.5$ and $1.0$) for varying activation profiles (linear (L) and non-linear (NL)), and for $A_0 = 0.1, 0.5$ and $1.0$ in the muscle belly. The distributions along the diagonal lines are shown in (a) and the lines along the muscle fiber direction are shown in (b).

Figure 6   Maximum principal strain ($\varepsilon_1$) distributions at three different simulation times ($t = 0.1, 0.5$ and $1.0$) for varying activation profiles (linear (L) and non-linear (NL)), and for $A_0 = 0.1, 0.5$ and $1.0$ in the muscle belly. The distributions along the diagonal lines are shown in (a) and the lines along the muscle fiber direction are shown in (b).

Figure 7   Maximum shear strain ($\gamma_{max}$) distributions at three different simulation times ($t = 0.1, 0.5$ and $1.0$) for varying activation profiles (linear (L) and non-linear (NL)), and for $A_0 = 0.1, 0.5$ and $1.0$ in the muscle belly. The distributions along the diagonal lines are shown in (a) and the lines along the muscle fiber direction are shown in (b).

linear (NL)) for $A_0 = 0.3$ and 0.7 in the muscle belly. The distributions along the diagonal lines are shown in (a) and the lines along the muscle fiber direction are shown in (b).

Figure 13  Maximum principal strain ($\varepsilon_1$) distributions at three different simulation times ($t = 0.1, 0.5$ and $1.0$) for varying activation profiles (linear (L) and non-linear (NL)) for $A_0 = 0.3$ and 0.7 in the muscle belly. The distributions along the diagonal lines are shown in (a) and the lines along the muscle fiber direction are shown in (b).

Figure 14  Maximum shear strain ($\gamma_{max}$) distributions at three different simulation times ($t = 0.1, 0.5$ and $1.0$) for varying activation profiles (linear (L) and non-linear (NL)) for $A_0 = 0.3$ and 0.7 in the muscle belly. The distributions along the diagonal lines are shown in (a) and the lines along the muscle fiber direction are shown in (b).

Figure 15  Evolution of the force output, pressure and maximum principal, maximum shear and volumetric strains (a), and their spearman correlation coefficients (b), for linear and non-linear activation profiles, for a maximal activation of "$A_0 = 0.3$", at locations $A - E$ on the muscle belly.

Figure 16  Evolution of the force output, pressure and maximum principal, maximum shear and volumetric strains (a), and their spearman correlation coefficients (b), for linear and non-linear activation profiles, for a maximal activation of "$A_0 = 0.7$", at locations $A - E$ on the muscle belly.

**Table Captions List**

Table 1      Material parameters of $W_\text{tendon}$ in Eq. (1) for tendon and aponeurosis (unit: N/cm²) [23].

Table 2      Material parameters of $W_\text{MT}^\text{dev}$ in Eq. (3) (unit: N/cm²).

Table 3      The deviatoric (and volumetric) relaxation and time coefficients from the five term Prony series for muscle tissue used in [27].

Table 4      The parameters of fifteen activation profiles for isometric contractions. $\bar{t}$ is the normalized time from 0 to 1.